\documentclass[preprint,12pt,authoryear]{elsarticle}

\usepackage{amssymb}
\usepackage[linkcolor=black,colorlinks=true,urlcolor=]{hyperref}
\usepackage{mathtools}

\usepackage{color,xcolor,ucs}
\usepackage{mathtools}   \usepackage{tikz} 
\usepackage{ amssymb }
\usepackage{extarrows} 
\usepackage{pgf,tikz}
\usepackage{float}
\usetikzlibrary{positioning}
\usetikzlibrary{shapes.geometric}
\usetikzlibrary{shapes.misc}
\usetikzlibrary{arrows}
\usepackage{caption}
\usepackage{mathrsfs}
\usetikzlibrary{arrows,shapes,automata,backgrounds,petri,positioning}
\usetikzlibrary{decorations.pathmorphing}
\usetikzlibrary{decorations.shapes}
\usetikzlibrary{decorations.text}
\usetikzlibrary{decorations.fractals}
\usetikzlibrary{decorations.footprints}
\usetikzlibrary{shadows}
\usetikzlibrary{calc}
\usetikzlibrary{spy}
\usepackage{amsmath}
\usepackage{array}
\usepackage{ amssymb }
\usepackage{braket}
\usepackage{qcircuit}
\usepackage{soul}
\usepackage{braket} 
\usepackage{relsize}

\usepackage{amsmath}
\usepackage{ amssymb }
\usepackage{braket}
\usepackage{qcircuit}
\usepackage{soul}
\usepackage{braket}
\usepackage{amsmath}

\journal{Information and Computation}

\begin{document}

\begin{frontmatter}



\title{Quantum Optimality in the Odd-Cycle game: the topological odd-blocker, marked connected components of the giant, consistency of pearls, vanishing homotopy} 


\author{Pete Rigas} 

\affiliation{ 
            city={Newport Beach},
            postcode={92625}, 
            state={CA},
            country={United States, pbr43@cornell.edu}}

\begin{abstract}
We characterize optimality of Quantum strategies for the Odd-Cycle game. Separate from other game-theoretic settings, parallel repetition for the Odd-Cycle game is related to the foam problem, which can be formulated through a minimization of the surface area. In comparison to previous works on minimizing the surface area, we quantify how properties of the marked giant connected component can be related to the maximum winning probability using Quantum strategies. Objects that we introduce to formulate such connections include the topological odd-blocker, previous examples of error bounds for other Quantum games that have been formulated by the author, pearls, consistent regions, and the cycle elimination problem. \footnote{\textbf{MSC Class}: 81P02; 81Q02}
\end{abstract}

\begin{graphicalabstract}
\end{graphicalabstract}

\begin{highlights}
\item We characterize optimality in the Odd-Cycle game through objects related to both Quantum information theory and topological combinatorics.
\item Fix $2 \leq d \leq 3$ and $n>0$. With previous results on the surface area of foams, which are equivalently related to Cycle Elimination, and Odd Cycle Elimination, problems, we characterize duality gaps between Classical and Quantum strategies through: (1) probabilistic estimates on the existence of foams, tubes and sections for which the diamond number is upper bounded, up to constants, by $n^d$; (2) comparing the probability of the previous item occurring with the probability of Alice and Bob winning with a restriction imposed on their strategies through a tensor contraction mapping; (3) establishing a similar comparison between the ratio of the two probabilities mentioned in the previous result under parallel repetition. The contraction mapping is defined in such a way that it allows (resp. prohibits) Alice and Bob from making use of a strategy unless the Quantum optimal value does (resp. does not) exceed a suitably chosen threshold when compared to the Classical optimal value.
\item Pursuing generalizations of maximizing the surface areas of foams under parallel repetition related to the arguments presented in this work is of interest.
\end{highlights}


\end{frontmatter}




\label{sec1}

\section{Introduction}

\subsection{Overview}

The Odd-Cycle game, discussed in [15], is a game-theoretic setting in which two players can make use of Quantum aspects of information to increase their probabilities of wining. To win such a game, the referee asks two players, Alice and Bob, to participate in the the following 2-coloring problem: each player is invited to color each vertex of a cycle independently with strictly positive probability $p$ with one color, and otherwise, with probability $1-p$, another color. Depending upon the total number of vertices present in a cycle, Alice and Bob can expect to win the game with a higher probability using Quantum, rather than classical, strategies. The vertices that Alice and Bob color along a cycle with a finite number of vertices is said to be \textit{two-colorable} if the players can convince the referee, with probability $1$, that the colors along each vertex of the cycle alternate (ie, that no two neighboring vertices have the same color). 

Quantum advantage has been realized for cycles with vertices $n$ satisfying $3 \leq n \leq 27$ [15]. The possibility of games that have previously been extensively characterized with classical strategies instead with Quantum strategies relates to several proposed quadratic, and exponential, speedups of Quantum algorithms and related paradigms [2, 3, 4, 5, 6, 7, 9, 10, 11]. Albeit the fact that Quantum advantage has been conjectured to hold in several information processing tasks, continuing to describe the fundamental workings of entanglement, and related paradoxical aspects, of Quantum information continues to remain of interest to explore [12, 13, 14, 16, 17, 19, 20, 21]. Besides industrial applications which could benefit from Quantum speedup, [22, 24, 25, 26, 27, 28, 29, 30, 31, 32, 33, 34], more theoretically driven aspects of Quantum information have recently emerged, ranging from: accelerated discovery of Quantum mechanical properties of materials [35, 36]; Quantum kernel methods for numerically approximating behaviors to a wide variety of nonlinear problems [38]; energy applications [39]; representations of probability distributions [40]; human cooperation, and statistical physics, [41]; generalized XOR games [43]; and several related directions of research [44, 47, 48, 49, 50].

Previous work of the author has sought to rigorously study all of the aforementioned aspects of Quantum Physics [44, 45, 46]. To further expand upon such aspects of Quantum Physics at the intersection of error bounds, optimality, and duality, we discuss how Quantum information can be used to the advantage of Alice and Bob for winning the Odd-Cycle game. With an analytical expression for the optimal value of the Odd-Cycle game, [15], error bounds, and related objects, can shed light upon proposed sources of Quantum advantage. Such sources of advantage ultimately reflect upon entanglement, which in the case of the Odd-Cycle game have been experimentally characterized using emerging photonic platforms [15]. Photonic experimental platforms differ from those that have been under significant development in industry, as qubits can be connected to each other using fiber optic cables operating at room temperature.

\bigskip

\noindent Before establishing connections with previous results of the author on optimality using Quantum strategies, we define several associated quantities for the Odd-Cycle game, including the set of players, action spaces, payoff functions, timing, and overall information structure that is evaluated against the referee's scoring predicate.

\bigskip

\noindent \textbf{Definition} \textit{1} (\textit{the set of players in the Odd-Cycle game}). Denote the set of players in the odd-Cycle game with $\big\{ \textit{Alice}, \textit{Bob} \big\}$.

\bigskip

\noindent Besides the structure of the number of players in the Odd-Cycle game, the manner in which Alice and Bob can make use of entanglement for prospective Quantum advantage is always of interest to explore, whether in a more theoretical or applied sense. As Alice and Bob prepare their responses for distribution to the referee, the scoring predicate ultimately determines whether the winning criteria is satisfied.

\bigskip

\noindent \textbf{Definition} \textit{2} (\textit{action spaces in the Odd-Cycle game}). Denote the action spaces of Alice and Bob with $\mathcal{A}$ and $\mathcal{B}$, respectively. From an iid sample of the referee's probability distribution $\pi$ of possible questions each action space takes the form,

\begin{align*}
      \mathcal{A} \equiv  \big\{   x \in \big\{ 0 , \cdots , n -1 \big\}    :   \textit{x is distributed to Alice from an iid sample $\sim \pi$}    \big\}  \\ \in \big\{ 0 , 1 \big\}      , \\ \\         \mathcal{B} \equiv \big\{  y \in \big\{ x , \big( x+1 \big) \mathrm{mod} n \big\}    :   \textit{y is distributed to Bob from iid sample $\sim \pi$}    \big\} \\ \in \big\{ 0 , 1 \big\}  . 
\end{align*}

\noindent \textbf{Definition} \textit{3} (\textit{outputs produced by Alice and Bob from their action spaces}). Denote $a \neq b \in \big\{ 0 , 1 \big\}$ corresponding to the outputs produced by Alice and Bob, respectively, from their sample spaces.

\bigskip

\noindent \textbf{Definition} \textit{4} (\textit{payoff functions in the Odd-Cycle game}). For the idd samples $\pi_1 \neq \pi_2 \sim \pi$ and outputs $a$ and $b$ from Alice and Bob, respectively, the payoff function is,

\[
 \left\{\!\begin{array}{ll@{}l} \underline{\textit{Payoff $+1$}}. \big\{ a = b \big\} \Longleftrightarrow \big\{  y = x \big\}  \\ \underline{\textit{Payoff $0$}}. \big\{  a \neq b \big\}  \Longleftrightarrow  \big\{             y = \big( x+1 \big) \mathrm{mod} n       \big\}        \end{array}\right.  \]

\bigskip

\noindent \textbf{Definition} \textit{5} (\textit{information structure of the Odd-Cycle game}). The information structure of the game consists of the following. At the beginning of each round Alice and Bob each only see their own input. Alice and Bob cannot communicate with each other after having received their input, however they can make use of entanglement to coordinate beforehand with $\rho_{AB}$.

\bigskip

\noindent \textbf{Definition} \textit{6} (\textit{timing of the Odd-Cycle game}). Before a round of the Odd-Cycle game begins, Alice and Bob agree on some strategy together which is dependent upon an entangled state that they share. After the referee samples $\pi_1 \neq \pi_2$ from his probability distribution, Alice and Bob independently, and without any communication, return their responses $a$ and $b$ to the referee. The referee, according to the above scoring predicate encoded in the payoff functions, determines whether the players receive payoffs $+1$ or $0$, respectively. The Odd-Cycle game can be repeated an arbitrarily many number of times for assessing the winning probabilities of Alice and Bob.

\bigskip

\noindent \textbf{Definition} \textit{7} (\textit{Classical and Quantum strategies in the Odd-Cycle game}). A Classical strategy that Alice and Bob can use for the Odd-Cycle game only depends upon shared randomness or deterministic rules. A Quantum strategy that Alice and Bob can use for the Odd-Cycle game depends upon entanglement that they can share through local Quantum measurements conditioned on the iid samples from the referee's probability distribution.

\bigskip

\noindent \textbf{Definition} \textit{8} (\textit{the referee's scoring predicate over Alice and Bob's strategies in the CHSH game}). Fix questions $x,y\sim \pi$, that is, the questions $x$ and $y$ drawn uniformly from the referee's probability distribution over questions, $\pi$, at random. If Alice receives $x$ and Bob receives $y$, the scoring predicate,

{\small \begin{align*}
  \mathscr{S}_A \big( x \big) \oplus \mathscr{S}_B \big( y \big) = \bigg[ \mathscr{S}_A \big( x \big) + \mathscr{S}_B \big( y \big) \bigg] \text{ } \mathrm{mod} 2 =                \left\{\!\begin{array}{ll@{}l} xy \Longleftrightarrow  \textit{Alice and Bob win the CHSH game,} \\ 0 \text{ } \textit{otherwise,} \end{array}\right.  , 
\end{align*} }

\noindent corresponds to the scoring criteria of the referee in the CHSH game over $\mathscr{T} \equiv \big( \mathscr{S}_A , \mathscr{S}_B \big)$, for,

\begin{align*}
   \mathscr{S}^d_A , \mathscr{S}^d_B : \big[ n \big]^d \longrightarrow \big[ 2 \big]^d , \\  \mathscr{S}_A , \mathscr{S}_B : \big[ n \big] \longrightarrow \big[ 2 \big] , 
\end{align*}

\noindent where,

\begin{align*}
  \mathscr{S}^d_A \big( x \big) = \mathscr{S}^d_B \big( x \big) = x \text{ } \mathrm{mod} 2  .
\end{align*}

\bigskip

\noindent \textbf{Definition} \textit{9} (\textit{reformulation of the CHSH scoring predicate in terms of the} $\delta$ \textit{functions}). One can reformulate the previously defined scoring predicate as,

{\small  \begin{align*}
          \mathcal{S}_A \big( x \big) \oplus \mathcal{S}_B \big( \widetilde{x+t} \big) \equiv  \big[ \mathcal{S}_A \big( x \big) +  \mathcal{S}_B \big( \widetilde{x+t} \big) \big] \text{ } \mathrm{mod}  2  =              \left\{\!\begin{array}{ll@{}l} t \oplus \delta \big( x , t \big)  \Longleftrightarrow  \textit{Alice and Bob win the} \\ \textit{CHSH game,} \\ 0 \text{ } \textit{otherwise,} \end{array}\right. 
                      , \\ 
\end{align*} }

\noindent where $\delta : \big[ n \big] \times \big\{ 0 , 1 \big\} \longrightarrow \big\{ 0 , 1 \big\}$, and,

\begin{align*}
   \underset{n, t \in \textbf{N}}{\sum} \delta \big( n , t \big) = \big( n +1 \big) \text{ } \mathrm{mod} 2 , \\    \underset{n, t \in \textbf{N}}{\sum} \delta \big( n , t \big) = 0  \text{ } \mathrm{mod} 2 .
\end{align*}

In comparison to Quantum states corresponding to optimal strategies of other games that have previously been characterized, [8, 23, 24, 37, 38, 42, 44, 46], the \textit{phase-shifted Bell state} for the Odd Cycle game is expressed as,

\begin{align*}
\ket{\psi_{\text{Odd-Cycle}}} \equiv       \frac{1}{\sqrt{2}}  \bigg[   \ket{0}_A \ket{1}_B + \mathrm{exp} \big( \theta \big)  \ket{1}_A \ket{0}_B        \bigg]    \text{, }
\end{align*}

\noindent for the rotation angle $\theta$ modulo $2\pi$ (ie, for $0 \leq \theta \leq 2 \pi$), and,

\begin{align*}
     \ket{0}_A \neq  \ket{1}_A ,  \ket{0}_A ,   \ket{1}_A  \in \textbf{C}^{d_A}  \text{, } \\   \ket{0}_B \neq  \ket{1}_B ,  \ket{0}_B , \ket{1}_B  \in \textbf{C}^{d_B}  \text{, }
\end{align*}

\noindent namely the possible responses of Quantum states in $0$, or $1$, from Alice and Bob's Hilbert spaces, respectively. In the field of Quantum Information theory, significant research have been devoted towards characterizing the action of classes of suitable operators from $\textbf{C}^{d_A}$ to $\textbf{C}^{d_B}$, and vice versa. For the Odd Cycle game, in comparison to other 2-player settings, ranging from the XOR, $\mathrm{XOR^{*}}$, and FFL, games, one encounters different barriers of achieving optimality using Quantum strategies, which is dependent upon the 2-colorability of a cycle. In comparison to the conditions that are enforced on the referee's predicate, through the scoring function $V$, for 2-player, and multiplayer, XOR games, error bounds, and various generalizations of error bounds, impose many conditions on potential sources of Quantum advantage.

To this end, mathematically characterizing the ways in which entanglement, and other paradoxical aspects of Quantum Information, inform proposed sources of Quantum advantage is of great interest. In the presence of entanglement, Quantum strategies corresponding to potential strategies of Alice and Bob for coloring vertices of the cycle could allow for winning probabilities of $1$, if there are an odd number of vertices in the cycle so that the colors of neighboring vertices always alternate. However, complications arise when players are independently coloring vertices of the cycle for an even number of vertices. By identifying strategies within a broader combinatorial space which can deter sources of Quantum advantage, several points of comparisons, and associations, between Odd Cycle, XOR, $\mathrm{XOR}^{*}$, and FFL, games, emerge.

\subsection{This paper's contributions}

\noindent This paper provides characterizations of error inequalities from the topology of cycles. Such cycles arise from the fact that, under two applications of parallel repetition, the optimal value of the Odd-Cycle game exhibits striking properties, relating to: (1) the existence of topological odd-cycles, which partition \textit{torical} graphs, ie graphs which block all odd cycles; (2) a partition of the space of strategies of Alice and Bob, for which the winning probability can be significantly increased; (3) the homotopy groups of oriented curves. Over the torus, whether a curve has vanishing homotopy group relates to a loop can be continuously contracted to a point. Topological properties of nonintersecting curves over \textit{torical} graphs determine whether Alice and Bob's strategy can satisfy the referee's scoring predicate. Albeit the fact that scoring predicates for the Odd-Cycle game share similarities to those for other games, several differences emerge. Predominantly, while parallel repetition for the Odd-Cycle game can be significantly increased, determining which strategies should not be included for Alice and Bob is dependent upon: (1) the probability that cycles of points on the \textit{torical} graph can be constructed; (2) whether consistent sets that are sufficiently large can be constructed over the \textit{torical} graph; (3) confirming that the homotopy group of paths within consistent sets is nonzero. From previous works that have previously been described within the Quantum Information theory literature, topological properties of cycles have not been thoroughly characterized. To further elaborate upon connections between two such fields, we quantify how error bounds can be adapted to reflect topological constraints on cycles.

To quantify how the ratio of winning probabilities,

\begin{align*}
\bigg\{  \frac{\textit{Alice and Bob's winning probability for the Odd-Cycle game using Quantu mstrategies}}{\textit{Alice and Bob's winning probability for the Odd-Cycle game using Classical strategies}} \bigg\}   ,
\end{align*}

\noindent behaves, we probabilistically relate the \textit{giant} of marked connected components to parallel repetition. Informally, while the collection of all marked connected components of the giant,

{\small \begin{align*}
  \mathcal{G} \equiv \underset{\textit{connected components}}{\bigcup}  \text{ } \text{ }  \overset{\cdot}{\underset{\textit{tube:tube} \cap \textbf{T}^2 \neq \emptyset}{\bigcup}}   \bigg\{ \forall \textit{connected components, } \exists \textit{tube} :  \textbf{P} \bigg[ \frac{\big| V \big( \textit{connected components} \big)\big| }{\big| V \big( \textit{tube} \big) \big|}    \\    \approx \frac{95}{100} \bigg] = 1     \bigg\}        , \\ 
\end{align*}} 

\noindent for tubes over the genus $1$ torus, $\textbf{T}^2$, introduced in \textbf{Definition} \textit{14}, given,

{\small \begin{align*}
  \big| V \big( \textit{connected components} \big)\big| \equiv \# \big\{  v \in V \big( \textbf{T}^2 \big) : v \textit{ has nonempty intersection with the vertex be-} \\ \textit{longing to a marked connected component contained within a tube over the torus}     \big\}              , \\ \\  \big| V \big( \textit{tube} \big) \big| \equiv   \# \big\{          v \in V \big( \textbf{T}^2 \big) : v \textit{ has nonempty intersection with a vertex belonging to the interior,} \\ \textit{ and also on the boundary, of a tube}  \big\}       . 
\end{align*} }

\bigskip

\noindent previous works in the literature have not articulated how the marked giant of connected components can be used to relate to foam problems. A \textit{foam problem}, as a computational problem that has been related to cycle, and odd-cycle, elimination problems discussed in [18], has previously been exploited in the context of parallel repetition. Remarkably, while parallel repetition for the Odd-Cycle game has been characterized topologically through the odd-blocker, other geometric aspects of parallel repetition emerge when formulating probabilistic quantities conditioned upon $\mathcal{G}$. However, before describing how parallel repetition for the Odd-Cycle game is unique in this regard, we draw the attention of the reader to previous works of the author which have characterized parallel repetition for two-player, and multiplayer, XOR games [44, 45]. The XOR game has received significant attention within the Quantum information theory for connections that it shares for the CHSH game. Moreover, it provides a source of prospective Quantum advantage for Alice and Bob, in the sense that they can win the game $11 \%$ more often using Quantum strategies, than with Classical strategies.

\subsection{Paper organization}

\noindent To establish connections between the topology of odd cycles and optimality of strategies in Quantum information theory, beginning in the next section we: (1) provide an overview of previous results of the author on two-player, and multiplayer, game-theoretic settings; (2) provide several characterization of optimality of strategies; (3) discuss the main results, which relate the existence of suitable topological odd-blockers to vanishing homotopy of cycles contained within consistent sets; (4) recapitulate topological objects introduced in [1, 18]. We conclude with arguments of each of the main results.

\subsection{Game-theoretic objects}

\subsubsection{Two-player games}

To investigate such directions of interest, we introduce several objects for 2-player, and multiplayer, XOR, $\mathrm{XOR}^{*}$, and FFL games. With the referee's predicate $V_{\text{Odd Cycle}}$, in place of the XOR predicate $V_{\mathrm{XOR}}$, inequalities corresponding to error bounds, and generalized error bounds, reflect upon the total number of optimal strategies. In the most simple XOR setting consisting of two players, given tensors that Alice and Bob can prepare for questions drawn from the referee's probability distribution, $i$ and $j$, with $A_i$ and $B_{ij}$, respectively, one error bound of interest takes the form, [37],

\begin{align*}
  \underset{1 \leq i < j \leq n}{\sum} \bigg\{ \text{ }  \bigg| \bigg|   \bigg[         \big( \frac{A_i + A_j}{\sqrt{2}} \big) \otimes \textbf{I} \bigg] \ket{\psi}    -   \big[  \textbf{I} \otimes B_{ij} \big] \ket{\psi}   \bigg|\bigg|^2 + \bigg| \bigg|   \bigg[ \big( \frac{A_i - A_j}{\sqrt{2}} \big)  \otimes \textbf{I}  \bigg] \ket{\psi}  \\   -   \big[  \textbf{I} \otimes B_{ji} \big]  \ket{\psi}     \bigg|\bigg|^2 \text{ }  \bigg\}  \text{, }
 \end{align*}

\noindent given a Quantum state 
$\ket{\psi}$ corresponding to the strategy of the two players. In the multiplayer setting, given the scoring predicate, which is constructed from a product probability distribution $\pi$ over $S \times T$, the game proceeds with the Referee examining the responses of Alice and Bob depending upon the entangled state that they share, in which, after sampling a pair $\big( S , T \big) \sim \pi$, and sending one question $s$ to Alice and another question $t$ to Bob,

\begin{align*}
    V \big(  s , t \big)   ab  \equiv 1 \Longleftrightarrow  \text{ Alice and Bob win,}    \\    V \big(  s , t \big)   ab \equiv -1 \Longleftrightarrow  \text{ Alice and Bob lose,}      
\end{align*}

\noindent with Alice's answer $a$, and Bob's answer $b$. Straightforwardly, the multiplayer predicate takes the above form, with the exception of the probability distribution for the referee's question being taken large enough so that a question can be distributed to each participant. Given the existence of a sufficiently small parameter $\epsilon$, in some game $G$ $\epsilon$-approximate optimality of the bias $\beta$, entails,

\begin{align*}
  \big( 1 - \epsilon \big) \beta \big( G \big)  \leq  \underset{\text{Questions}}{\sum} \bra{\text{Optimal Strategy}} \bigg[    \underset{\text{Players } i}{\bigotimes}   \text{Tensors of player observables}   \bigg]  \\ \times   \ket{\text{Optimal Strategy}}  \leq      \beta \big( G \big)   \text{, }
\end{align*}

\noindent for the supremum of the bias,

\begin{align*}
 \beta \big( G \big) \equiv  \underset{\text{Strategies }\mathcal{S}}{\mathrm{sup}} \text{ } \beta \big( G , \mathcal{S} \big) \text{, } 
\end{align*}

\noindent over the combinatorial set $\mathcal{S}$ of all possible strategies. In the multiplayer setting, previous work of the author, [46], characterized more complicated manners in which entanglement can arise. That is, in comparison to the Bell states, or equivalently, the EPR pairs, arising through entanglement in tensor products of two operators in the $n\equiv 2$ Bell states for the CHSH$\big( n\big)$ game,

\begin{align*}
\bigg[  \textbf{I} \otimes \textbf{I}  \bigg] \bigg[  \frac{\ket{00} + \ket{11}}{\sqrt{2}}  \bigg] = \frac{\ket{00} + \ket{11}}{\sqrt{2}}   \text{ } \text{ , }     \bigg[     \sigma_x \otimes \textbf{I}  \bigg] \bigg[ \frac{\ket{00} + \ket{11}}{\sqrt{2}} \bigg]   = \frac{\ket{10} + \ket{01}}{\sqrt{2}}   \text{, }   \\   \bigg[   \sigma_z    \otimes \textbf{I}  \bigg]  \bigg[ \frac{\ket{00} + \ket{11}}{\sqrt{2}} \bigg]  = \frac{\ket{00} - \ket{11}}{\sqrt{2}}       \text{ } \text{ , }    \bigg[ \sigma_x \sigma_z     \otimes   \textbf{I}   \bigg] \bigg[ \frac{\ket{00}+\ket{11}}{\sqrt{2}} \bigg]  = \frac{\ket{10}- \ket{01}}{\sqrt{2}}     \text{, } 
\end{align*}

\noindent higher-dimensional analogs of entanglement arise from operations of the form operators of the form, [46],

\begin{align*}
 \bigg[ \frac{\sigma_z \otimes \textbf{I} \otimes \overset{N-3}{\cdots} \otimes \textbf{I}}{\sqrt{N}} \bigg]   \bigg[    \underset{1 \leq j \leq N}{\sum} \ket{\text{Player } j \text{ state}}      \bigg] \text{, } \end{align*}

\noindent for $N>0$ players. In the Odd-Cycle game, given the existence of the optimal strategy for Alice and Bob that is encoded with $\ket{\psi_{\text{Odd Cycle}}}$, one would expect to consideration the following summation,

\begin{align*}
   \bra{\psi_{\text{Odd Cycle}}}  A_s \otimes B_t      \ket{\psi_{\text{Odd Cycle}}} \equiv \bigg[  \frac{1}{\sqrt{2}}  \bigg[   \bra{0}_A \bra{1}_B + \mathrm{exp} \big( \theta \big)  \bra{1}_A \bra{0}_B        \bigg] \bigg]   \bigg[  A_s \otimes B_t       \bigg]  \\ \times \bigg[   \frac{1}{\sqrt{2}}  \bigg[   \ket{0}_A \ket{1}_B   + \mathrm{exp} \big( \theta \big)  \ket{1}_A \ket{0}_B        \bigg]   \bigg] \end{align*}

   \begin{align*} =      \bigg[ \frac{1}{\sqrt{2}} \bigg]^2     \bigg\{  \bra{0}_A \big[ \bra{1}_B A_s \otimes B_t \ket{0}_A \big]  \ket{1}_B  +          \mathrm{exp} \big( \theta \big)^2    \bra{1}_A \big[ \bra{0}_B A_s \otimes B_t \ket{1}_A  \big]     \ket{0}_B     \bigg\}  \\ \\  =   \bigg[ \frac{1}{\sqrt{2}} \bigg]^2     \bigg\{  \bra{01}_{AB}  A_s \otimes B_t \ket{01}_{AB} +          \mathrm{exp} \big( \theta \big)^2    \bra{10}_{AB}  A_s \otimes B_t \ket{10}_{AB}        \bigg\}       \text{, }
\end{align*}

\noindent of expectation values. For various quantitative characterizations of optimality, and near optimality, this reflects upon the XOR optimal value, which is given by,

\begin{align*}
 \frac{1}{\sqrt{2}} \equiv  \underset{A_i , B_{jk} , \psi}{\mathrm{sup}} \text{ }  \frac{1}{4 {n \choose 2}}   \underset{1\leq i \leq j \leq n}{\sum}  \bra{\psi} \bigg[       A_i B_{ij} +  A_j B_{ij}  +   A_i B_{ji}  -   A_j B_{ji}  \bigg]           \ket{\psi} 
               \text{, } 
\end{align*}

\noindent for the optimal $2$-$\mathrm{XOR}$ strategy, $\ket{\psi_{2\mathrm{XOR}}} \equiv  \ket{\psi}$. Hence, given the expression for the state, $\bra{\psi_{\text{Odd Cycle}}}  A_s \otimes B_t      \ket{\psi_{\text{Odd Cycle}}}$, in the Odd-Cycle game, it is of interest to determine how the optimal value, equivalently the maximum winning probability, of the game depends upon the optimal strategy. For $2$-$\mathrm{XOR}$, and multiplayer XOR, games alike, $\epsilon$ approximality, for $\epsilon$ taken to be sufficiently small, of exact and approximate strategies implies $\epsilon$ approximality of the bias $\beta \big( G \big)$. Explicitly, the bias takes the form, for $\ket{\psi} \in \textbf{C}^{d_A} \otimes \textbf{C}^{d_B}$ and the space of possible strategies $\mathcal{S}$,

\begin{align*}
      \beta \big( G , \mathcal{S}  \big)     \equiv \underset{s \in S}{\sum} \underset{t \in T}{\sum} G_{st} \bra{\psi} A_s \otimes B_t \ket{\psi}  \text{, } 
\end{align*}

\noindent for questions $s$ and $t$ that are respectively distributed to Alice and Bob, from the set of questions $S$ and $T$ belonging to the referee's probability distribution $\pi$, while the $\epsilon$-approximality condition takes the form,

\begin{align*}
     \big( 1 - \epsilon \big)   \underset{\mathcal{S}}{\mathrm{sup}} \text{ } \beta \big( G , \mathcal{S} \big)   \equiv \big( 1 - \epsilon \big) \beta \big( G \big)   \leq   \beta \big( G , \mathcal{S} \big)        \leq  \beta \big( G \big)         \equiv  \underset{\mathcal{S}}{\mathrm{sup}} \text{ } \beta \big( G , \mathcal{S} \big)      \text{, }
\end{align*}

\noindent from the fact that,

\begin{align*}
    \frac{1}{\sqrt{2}} \big( 1 - \epsilon \big) \leq    \underset{A_i , B_{jk} , \psi}{\mathrm{sup}}   \frac{1}{4 {n \choose 2}} \bigg\{   \underset{1\leq i \leq j \leq n}{\sum}  \bra{\psi} \bigg[       A_i B_{ij} +  A_j B_{ij}  +   A_i B_{ji}  -   A_j B_{ji}  \bigg]           \ket{\psi} \bigg\}  \leq \frac{1}{\sqrt{2}}          \text{, } 
\end{align*}

\noindent for $2$ players. Beyond the $2$-$\mathrm{XOR}$ game, one encounters a rich information-theoretic landscape, which allows for the possibility of: identifying barriers to optimality using Quantum strategies that can depend on entanglement, in comparison to those that have been classified with classical strategies; establishing connections between whether Quantum optimality for games can be obtained depending upon the referee's scoring predicate; formulating connections between Game theory and extremal combinatorics; establishing rates of decay of ordinary, and strong, parallel repetition of optimal values; amongst several other directions of research.

At the crossroads of all such areas, the Odd-Cycle game is of interest to explore, not only from the fact that the referee's scoring predicate depends upon 2-colorability of cycles, but also from the fact that responses from Alice and Bob can be analyzed from error bounds for XOR, $\mathrm{XOR}^{*}$, and FFL, games [37, 44, 46]. Quantum mechanically, the error bound for games that players can interact with using strategies other than classical one quantifies near, and approximate, optimality. While error bounds, and generalizations of error bounds, provide information-theoretic consequences on potential advantages of Quantum information, barriers to obtaining such advantages persist. Generally, if the Quantum value of a game $G$ is identically $1$, namely,

\begin{align*}
  \omega_q \big( G \big) \equiv 1  \text{, }
\end{align*}

\noindent then one can achieve the highest level of proposed Quantum advantage, as in the magic square game, which implies that pseudo-telepathy can be achieved. In the Odd-Cycle game, despite the fact that,

\begin{align*}
  \omega_q \big( \text{Odd-Cycle} \big) < 1  \text{, }
\end{align*}

\noindent proposed sources of Quantum advantage still exist, which can help Alice and Bob as they are determining which colors should be used on each vertex of the cycle.

\subsubsection{Overview of previous game-theoretic results}

\noindent We state the collection of results from the 2-player setting which depend on the optimal value, and then discuss how exponential rates of decay for parallel repetition of the optimal value come into play.

Denote the Frobenius norm,

\begin{align*}
  \big|\big| A \big|\big|_F \equiv \sqrt{\overset{m}{\underset{i=1}{\sum}} \overset{n}{\underset{j=1}{\sum}} \big| a_{ij} \big|^2 } = \sqrt{\mathrm{Tr} \big[ A^{\dagger} A \big] }  \text{, } 
\end{align*}

\noindent of an $m \times n$ matrix $A$ with entries $a_{ij}$, there exists a \textit{linear bijection} $\mathcal{L}$ between the tensor product space, $\textbf{C}^{d_A} \otimes \textbf{C}^{d_B}$, and the space of $d_A \times d_B$ matrices with complex entries, $\mathrm{Mat}_{d_A , d_B} \big( \textbf{C} \big)$, satisfying (\textbf{Lemma} \textit{1}, {[23]}),

\begin{itemize}
\item[$\bullet$] \underline{\textit{Image of the tensor product of two Quantum states under} $\mathcal{L}$}: $\forall \ket{u} \in \textbf{C}^{d_A}, \ket{w} \in \textbf{C}^{d_B}, \exists \ket{u^{*}} \in \textbf{C}^{d_B} : \mathcal{L} \big( \ket{u} \otimes \ket{w} \big) = \ket{u} \bra{u^{*}}  \text{, }$ 
\item[$\bullet$] \underline{\textit{Product of a matrix with the image of a Quantum state under} $\mathcal{L}$}: $\forall \ket{u} \in \textbf{C}^{d_A}, \exists A \in \mathrm{Mat}_{d_A} \big( \textbf{C} \big) : A \mathcal{L} \big( \ket{u} \big) = \mathcal{L} \big( A \otimes I \ket{u} \big)\text{, }$
\item[$\bullet$] \underline{\textit{Product of the image of a Quantum state under $\mathcal{L}$ with the transpose of a matrix}}:  $\forall \ket{w} \in \textbf{C}^{d_B}, \exists B \in \mathrm{Mat}_{d_B} \big( \textbf{C} \big) : \mathcal{L} \big( \ket{w} \big) B^T = \mathcal{L} \big( I \otimes B \ket{w} \big)  \text{, }$
\item[$\bullet$] \underline{\textit{Frobenius norm equality}}: $\forall \ket{w} \in \textbf{C}^{d_B} : \big|\big| \mathcal{L} \big(   \ket{w}     \big) 
 \big|\big|_F = \ket{w}  \text{. } $
\end{itemize}

\bigskip

\noindent \textbf{Lemma} \textit{5B}, [45] (\textit{strong parallel repetition of }$\sqrt{\epsilon^{\wedge}}$- \textit{FFL approximality}, \textbf{Lemma} \textit{8}, [44]). From the same quantities introduced in the previous result, one has,

\begin{align*}
   \bigg| \bigg|               \bigg[ \big(  A_k  \wedge A_{k^{\prime}} \big) \otimes \textbf{I} \bigg] \ket{\psi_{\mathrm{FFL} \wedge \mathrm{FFL}}}    -  \bigg[ \textbf{I} \otimes \bigg[     \frac{\pm \big(  B_{kl} \wedge B_{k^{\prime} l^{\prime}} \big)  + \big( B_{lk} \wedge B_{k^{\prime} l^{\prime}} \big) }{\big| \pm \big(  B_{kl} \wedge B_{k^{\prime} l^{\prime}} \big)  + \big( B_{lk} \wedge B_{k^{\prime} l^{\prime}} \big) \big| }           \bigg] \bigg] \\ \times \ket{\psi_{\mathrm{FFL} \wedge \mathrm{FFL}}}              \bigg| \bigg|    < 20 \sqrt{N \epsilon^{\wedge}}     \text{. } 
\end{align*}

\noindent The results from previous work of the author above, originally stated in [46], are of great interest to adapt, and further explore, for the Odd-Cycle game. Intuitively, exactly optimal strategies for the Odd-Cycle game with Quantum strategies would entail that Alice and Bob would have perfect knowledge of the color of each vertex along the cycle. Relatedly, approximate optimality using Quantum strategies for the Odd-Cycle game would entail that there could be some probability, sufficiently small, so that Alice and Bob color neighboring vertices in the cycle with the same color, hence resulting in a vanishing of the referee's scoring predicate, namely a configuration in which Alice and Bob would not win the Odd-Cycle game.

Given the aforementioned differences between the referee's predicate for the XOR, and Odd-Cycle, games, in the upcoming subsections we discuss how exact, and approximate, optimality can be analyzed. Barriers to optimality using Quantum, in comparison to Classical, strategies for the Odd-Cycle game capture different ways in which entanglement can influence the decisions of Alice and Bob.

\subsection{Main results}

\subsubsection{Motivation}

\noindent We provide a motivation of topological objects that will be connected to error bounds and related game-theoretic objects introduced in the previous section, relating to the discussion provided in [18]. That is, introduce a $\textbf{Z}^2$ \textit{foam} $F$, where,

\begin{align*}
 F \in  F \big( \textbf{Z}^2 \big)       , 
\end{align*}

\noindent whose supremum of the surface area,

\begin{align*}
    \underset{F \in F ( \textbf{Z}^2 )}{\mathrm{inf}} \big\{ \textit{Surface area of a foam } F \big\} 
\end{align*}

\noindent is realized through the spanning set,

\begin{align*}
     \underset{\textbf{R}^2}{\mathrm{span}} \big\{ \textit{hexagons with two 120 degree angles} \big\}      ,
\end{align*}

\noindent corresponding to a hexagonal tiling of $\textbf{Z}^2$. The surjection,

\begin{align*}
    \varphi : \textbf{Z}^2 \longrightarrow \textbf{T}^2 \mapsto \textit{Two-dimensional torus} \equiv G \equiv \big( V \big( \textbf{T}^2 \big) , E \big( \textbf{T}^2 \big) \big) , 
\end{align*}

\noindent allows one to periodically embed the above set of linear combinations over $\textbf{R}^2$ to over $\textbf{T}^2$. Straightforwardly, the spanning set for maximizing the surface area of the foam that is periodically embedded over $\textbf{T}^2$ instead takes the form,

\begin{align*}
     \underset{\textbf{T}^2}{\mathrm{span}} \big\{ \textit{hexagons with two 120 degree angles} \big\}      .
\end{align*}

\noindent For $d\equiv 2$ corresponding to \textit{two} applications of ordinary parallel repetition, the ordinary, and odd, cycle \textit{elimination problems} quantify the asymptotic behavior of,

\begin{align*}
  \underset{\textbf{T}^2}{\textbf{E}} \big[  \# \big\{ \textit{edges} :  \textit{edges in cubes, and sections, which have nonempty intersection with a tube} \big\} \big]  , 
\end{align*}

\noindent where the number of sections, tubes, and cubes, are given by,

\begin{align*}
    \textit{Sections } \mathcal{S} \equiv \underset{S \in \mathcal{S}}{\bigcup} \big\{ S: S \textit{ intersects finitely many faces of a hexagonal tiling} \big\}   , \end{align*}
    
    \begin{align*} \textit{Tubes } \mathcal{T}  \equiv \underset{T \in \mathcal{T}}{\bigcup} \big\{    T: T \textit{ intersects finitely many sections}   \big\}  , \end{align*}
    
    \begin{align*} \textit{Cubes } \mathcal{C} \equiv \underset{C \in \mathcal{C}}{\bigcup} \big\{ C : C \textit{ intersects finitely many tubes} \big\} , 
\end{align*}

\noindent respectively. With the following above topological objects, formally, introduce the following probabilistic quantities:

\bigskip

\noindent \textbf{Definition} \textit{10} (\textit{Cycle elimination problem over the torus}). Denote the vertex set over $\textbf{Z}^d$ as,

\begin{align*}
V \big( \textbf{Z}^d \big)   \equiv \big\{ \textit{vertices } v_1 \neq v_2  : L_{\infty} \big\{ v_1 , v_2 \big\}   = 1 \big\}   ,
\end{align*}

\noindent for the $L$-$\infty$ norm, $L_{\infty} \big\{ \cdot , \cdot \big\} $, along with the edge set,

\begin{align*}
  E \big( \textbf{Z}^d \big) \equiv \big\{ \textit{edges } e_1, e_2 : \big| e_1 - e_2 \big| = 1 \big\}   ,
\end{align*}

\noindent from which one can introduce,

{\small \begin{align*}
   \mathscr{C}\mathscr{E} \mathscr{P} \equiv \underset{\textit{vertices} \in V ( \textbf{Z}^d ) }{\bigcup} \bigg\{  \forall  \textit{vertices} ,  \exists e \in E \big( \textbf{Z}^d \big)  , \delta > 0   :   \big\{   \textit{probability that there exists a topological} \\ \textit{  cycle without removing countably many vertices}   > 0        \big\} , \big\{   \textit{probability that there exists an} \\ \textit{ topological Cycle after having removed a } \delta \textit{ fraction of vertices from the} \\ \textit{ cycle} =  0        \big\}     \bigg\}       , 
\end{align*} }

\noindent corresponding to the \textit{cycle elimination problem} over the $d$-dimensional torus, $\textbf{T}^d$.

\bigskip

\noindent \textbf{Definition} \textit{11} (\textit{Odd-cycle elimination problem over the torus}). With $V \big( \textbf{Z}^d \big)$ and $E \big( \textbf{Z}^d \big)$ defined in \textbf{Definition} \textit{1}, introduce,

{\small \begin{align*}
   \mathscr{O}\mathscr{C}\mathscr{E} \mathscr{P} \equiv \underset{\textit{vertices} \in V ( \textbf{Z}^d ) }{\bigcup} \bigg\{  \forall  \textit{vertices}  ,  \exists e \in E \big( \textbf{Z}^d \big)  , \delta^{\prime} > 0   :   \big\{   \textit{probability that there exists an} \\ \textit{  Odd-Cycle without removing countably many vertices}   > 0        \big\} , \big\{   \textit{probability that there exists} \\ \textit{ an Odd-Cycle  after having removed a } \delta^{\prime} \textit{ fraction of vertices from the} \\ \textit{cycle} =  0        \big\}     \bigg\}       , 
\end{align*} }

\noindent corresponding to the \textit{Odd-Cycle elimination problem} over the $d$-dimensional torus, $\textbf{T}^d$.

\bigskip

\noindent \textbf{Definition} \textit{12} (\textit{Odd-Cycle parallel repetition problem}). Fix $m$ odd and $d \leq m$. Under the assumption that the optimal value of winning the $m$ Odd-Cycle game takes the form,

\begin{align*}
   \mathrm{Val} \big( \textit{m Odd-Cycle game} \big) \equiv  \omega_q \big( \text{Odd-Cycle} \big) = 1 - \Theta \big( m^{-1} \big)  ,
\end{align*}

\noindent does there exist an expression for parallel repetition of the optimal value,

\begin{align*}
   \mathrm{Val} \big( \big(  \textit{m Odd-Cycle game} \big)^{\otimes d} \big) \equiv  \omega_q \big( \text{Odd-Cycle}^{\otimes d} \big) = \omega_q \big( \text{Odd-Cycle} \big)^d ,
\end{align*}

\noindent under $d$ rounds of parallel repetition, which takes the following form,

\begin{align*}
   \mathrm{Val} \big( \big(  \textit{m Odd-Cycle game} \big)^{\otimes d} \big) \leq 1 - m^{-1} \Omega \big( d \big)   ?
\end{align*}

\noindent Several problems relating to the parallel repetition problem have been previously examined by the author.

\bigskip

\noindent \textbf{Definition} \textit{13} (\textit{Foam problem over quotient spaces}). Consider the quotient space $\textbf{R}^d \backslash \textbf{Z}^d$. If unique, does there exist a quotient minimizer, $\mathscr{M}$, where,

\begin{align*}
\mathscr{M} \equiv  \underset{d}{\mathrm{inf}} \big\{     A \big( d \big) : \textit{the surface area A over } \textbf{R}^d  \textit{ is tiled by } \textbf{Z}^d   \big\}    ?
\end{align*}

\bigskip

\noindent \textbf{Definition} \textit{14} (\textit{Degree of a section}). Introduce,

\begin{align*}
  \mathrm{deg} \big( \mathcal{S} \big) \equiv d \big( \mathcal{S} \big) = \underset{x \in \mathcal{S}}{\bigcup} \big| \big\{    x : x \in S , S \in \mathcal{S}       \big\} \big|   ,
\end{align*}

\noindent corresponding to the degree of $\mathcal{S}$.

\bigskip

\noindent \textbf{Definition} \textit{15} (\textit{Distributions over sections, cubes, and tubes}). Introduce,

\begin{align*}
 \mathcal{D}_{\mathcal{S}}  \equiv       \underset{S \in \mathcal{S}}{\bigcup} \big\{ \textit{Distributions over marked points in a section S} \big\}    =  \underset{S \in \mathcal{S}}{\bigcup} \mathcal{D}_S  ,
\end{align*}

\noindent corresponding to distributions over sections. Straightforwardly, one can define $\mathcal{D}_{\mathcal{C}}$ and $\mathcal{D}_{\mathcal{T}}$ in the same way.

\bigskip

\noindent \textbf{Definition} \textit{16} (\textit{Relative distributions over sections, cubes, and tubes}). Introduce,

\begin{align*}
  \mathcal{D}_{\mathrm{Rel } \text{ } \mathcal{S}}  \equiv       \underset{S \in \mathcal{S}}{\bigcup} \big\{  \big|  \big\{ \textit{marked points} \big\}  \cap S   \big| \big| S \big|^{-1}  \big\}          ,
\end{align*}

\noindent corresponding to the relative distribution over sections. Straightforwardly, one can define the relative distributions for cubes and tubes in the same way.

\bigskip

\noindent In the next subsection, we further describe how the foam problem over $\textbf{R}^d$ can be quantified with respect to the diamond, and $L_{+\infty}$, norms.

\subsubsection{Discrete probabilistic objects in the main results}

\noindent We provide a statement of the main results for the Odd-Cycle game.

\bigskip

\noindent \textbf{Definition} \textit{17} (\textit{discrete probabilistic objects with respect to the L}-$\infty$, \textit{and diamond norms}). Introduce the probabilities,

\begin{align*}
 \mathcal{P}_1 \equiv   \textbf{P} \bigg[   \forall n \equiv 2, \exists d \equiv 2: \underset{\textit{Surface Area}}{\mathrm{inf}}  \big\{ \mathrm{span} \big\{ e_{1,d}, e_{2,d} \big\} \big\}    \lesssim n^d                \bigg]      ,
\end{align*}

\noindent corresponding to the probability that the maximum surface area of a foam is up to constant upper bounded by $n^d$,

\begin{align*}
    \mathcal{P}_2 \equiv     \textbf{P} \bigg[   \forall n \equiv 2, \exists d \equiv 2:     \underset{\textit{Surface Area}}{\mathrm{inf}}  \big\{    L_{\infty}         \big\{    e_{1,d} , e_{2,d}   \big\}                 \big\}   \lesssim n^d                    \bigg]        ,
\end{align*}

\noindent corresponding to the probability that , for the $L$-$\infty$ norm, $L_{+\infty} \big\{ \cdot , \cdot \big\} $,

\begin{align*}
  \mathcal{P}_3 \equiv   \textbf{P} \bigg[   \forall n \equiv 2, \exists d \equiv 2:        \underset{\textit{Surface Area}}{\mathrm{inf}}  \big\{    d_{\textit{Diamond}}         \big\{ e_{1,d} , e_{2,d}  \big\}        \big\}     \lesssim n^d          \bigg]          ,
\end{align*}

\noindent corresponding to the probability that the supremum of diamond norms $d_{\textit{diamond}} \big\{ \cdot , \cdot \big\} $ between the basis elements $e_{1,d}$ and $e_{2,d}$ is of order $n^d$, given the diamond norm,

  \begin{align*}
           d_{\textit{Diamond}} \big\{  \Phi  , X \big\}   \equiv  \big| \big|  \big( \Phi \otimes \textbf{I}_n \big) X \big| \big|_1      , 
        \end{align*}

\noindent for the mapping $\Phi : \textbf{M}_n \big( \textbf{C} \big) \longrightarrow \textbf{M}_m \big( \textbf{C}\big)$, namely the collection of $m\times n$ matrices with elements over the base field $\textbf{C}$, $X \in \textbf{M}_{n^2} \big( \textbf{C} \big)$ and the identity map $\textbf{I}_n : \textbf{M}_n \big( \textbf{C} \big) \longrightarrow \textbf{M}_n \big( \textbf{C} \big)$. Related to the three probabilities above, also introduce,

\begin{align*}
 \mathcal{P}_4 \equiv    \textbf{P} \bigg[   \forall n \equiv 2, \exists d \equiv 2:         d_{\textit{Diamond}}      \bigg\{  \underset{\textit{Surface Area}}{\mathrm{inf}}     \big\{ e_{1,d} , e_{2,d}  \big\}        \bigg\}     \lesssim n^d          \bigg]          ,
\end{align*}

\noindent corresponding to the probability that the diamond norm of the large difference between $e_{1,d}$ and $e_{2,d}$ is of order $n^d$. One can similarly define a spanning set,

\begin{align*}
  \mathrm{span} \big\{ e_{1,d} , e_{2,d} , e_{3,d} \big\}   ,
\end{align*}

\noindent for $d\equiv 3$, as well as the minimum,

{\small \begin{align*}
   L_{\infty}  \bigg\{                  \underset{d}{\mathrm{min}}             \big\{   \big\{ e_{1,d} , e_{2,d} \big\} , \big\{ e_{1,d} , e_{3,d}  \big\} , \big\{ e_{2,d} , e_{3,d} \big\}        \big\}              \bigg\}   \equiv  L_{\infty}  \bigg\{                  \underset{d}{\mathrm{min}}      \big\{ e_{1,d} , e_{2,d} \big\} ,    \underset{d}{\mathrm{min}}      \big\{ e_{2,d} , e_{3,d} \big\}       \bigg\}          ,
\end{align*} }

\noindent with respect to $L_{\infty} \big\{ \cdot , \cdot \big\} $, in addition to the minimum,

{\small \begin{align*}
 d_{\textit{Diamond}} \bigg\{                  \underset{d}{\mathrm{min}}             \big\{   \big\{ e_{1,d} , e_{2,d} \big\} , \big\{ e_{1,d} , e_{3,d}  \big\} , \big\{ e_{2,d} , e_{3,d} \big\}        \big\}              \bigg\}    \equiv   d_{\textit{Diamond}}    \bigg\{   \underset{d}{\mathrm{min}}      \big\{ e_{1,d} , e_{2,d} \big\} \\ ,    \underset{d}{\mathrm{min}}      \big\{ e_{2,d} , e_{3,d} \big\}         \bigg\}    ,
\end{align*} }

\noindent with respect to $d_{\textit{Diamond}} \big\{ \cdot , \cdot \big\} $. 

\bigskip

\noindent To demonstrate that a correspondence of the form,

\begin{align*}
 \bigg\{  \big\{  \mathcal{P}_1 > 0 \big\} \Longleftrightarrow   \big\{  \mathcal{P}_2 > 0 \big\} \bigg\} \Longleftrightarrow  \bigg\{  \big\{  \mathcal{P}_3 > 0 \big\}  \Longleftrightarrow   \big\{  \mathcal{P}_4 > 0 \big\} \bigg\}  ,
\end{align*}

\noindent holds, it suffices to perform a computation of a quantity of the form,

\begin{align*}
   \mathcal{P}_5 \equiv   \textbf{P}    \bigg[                \forall n \equiv 2 , \exists d \equiv 3 :   d_{\textit{Diamond}}    \bigg\{   \underset{d}{\mathrm{min}}      \big\{ e_{1,d} , e_{2,d} \big\} ,    \underset{d}{\mathrm{min}}      \big\{ e_{2,d} , e_{3,d} \big\}         \bigg\}      \lesssim n^d                      \bigg]          ,
\end{align*}

\noindent corresponding to the probability that the surface area of a three-dimensional foam is of order $n^d$ after two rounds of ordinary parallel repetition. While we do not fully explore the behavior of the above probability in this work, an adaptation of the forthcoming arguments for one round of parallel repetition, in addition to other arguments, could be of interest to explore in the future.

\bigskip

\noindent Given the previous expected correspondence between the probabilities $\mathcal{P}_1, \cdots , \mathcal{P}_4$, one would also expect,

\begin{align*}
  \big\{ \mathcal{P}_1 > 0 \big\} \Longleftrightarrow     \big\{ \mathcal{P}_5 > 0 \big\}   .
    \end{align*}

\noindent With respect to the number oof ordinary parallel repetition operations, observe,

\begin{align*}
    \mathcal{P}_5 \propto  \textbf{P} \bigg[ \forall n \equiv 2 , \exists  1 \leq d^{\prime} \leq d :  \underset{d^{\prime}}{\mathrm{sup}}             \bigg\{  d_{\textit{Diamond}}  \big\{ e_{1,d^{\prime}} , e_{2,d^{\prime}}  \big\} ,         d_{\textit{Diamond}}  \big\{ e_{1,d^{\prime}} , e_{3,d^{\prime}}  \big\} \\ , d_{\textit{Diamond}}  \big\{ e_{2,d^{\prime}}  , e_{3,d^{\prime}}  \big\}       \bigg\}          \lesssim n^d                   \bigg]                       .
\end{align*}

\noindent The above probability can equal any one of the following three expressions, 

\begin{align*}
  \underset{d \equiv 3}{\mathrm{sup}} \text{ }  \textbf{P} \bigg[  \forall   n \equiv 2  , \exists 1 \leq d^{\prime}  \leq d     :       d_{\textit{Diamond}} \big\{ e_{1,d^{\prime}} , e_{3, d^{\prime}} \big\} \Longleftrightarrow d_{\textit{Diamond}}    \bigg\{      \big\{ e_{1,d^{\prime}} , e_{3,d^{\prime}} \big\}    \textbf{1}_1                      \bigg\}      \bigg]     , \\ \\  \underset{d \equiv 3}{\mathrm{sup}} \text{ }  \textbf{P} \bigg[  \forall   n \equiv 2  , \exists 1 \leq d^{\prime}  \leq d     :       d_{\textit{Diamond}} \big\{ e_{1,d^{\prime}} , e_{2, d^{\prime}} \big\} \Longleftrightarrow d_{\textit{Diamond}}    \bigg\{      \big\{ e_{1,d^{\prime}} , e_{2,d^{\prime}} \big\}    \textbf{1}_2                      \bigg\}      \bigg]  , \\ \\  \underset{d \equiv 3}{\mathrm{sup}} \text{ }  \textbf{P} \bigg[  \forall   n \equiv 2  , \exists 1 \leq d^{\prime}  \leq d     :       d_{\textit{Diamond}} \big\{ e_{2,d^{\prime}} , e_{3, d^{\prime}} \big\} \Longleftrightarrow d_{\textit{Diamond}}    \bigg\{      \big\{ e_{2,d^{\prime}} , e_{3,d^{\prime}} \big\}    \textbf{1}_3                      \bigg\}      \bigg]  , 
\end{align*}

\noindent where the three indicator functions take the form,

\begin{align*}
 \textbf{1}_1     \equiv \textbf{1}       \bigg\{      \underset{1 \leq d^{\prime} \leq d }{\mathrm{min}} \big\{  \big\{ e_{1,d^{\prime}}  , e_{2,d^{\prime}} \big\} ,   \big\{         e_{1,d^{\prime}} , e_{3,d^{\prime}} \big\} , \big\{ e_{2,d^{\prime}} , e_{3,d^{\prime}} \big\} \big\}    \neq        \big\{ e_{1,d^{\prime}} , e_{2,d^{\prime}} \big\}  ,  \big\{ e_{2,d^{\prime}} , e_{3,d^{\prime}}    \big\}    \bigg\}          , 
\\ \\ 
 \textbf{1}_2     \equiv \textbf{1}       \bigg\{      \underset{1 \leq d^{\prime} \leq d }{\mathrm{min}} \big\{  \big\{ e_{1,d^{\prime}}  , e_{2,d^{\prime}} \big\} ,   \big\{         e_{1,d^{\prime}} , e_{3,d^{\prime}} \big\} , \big\{ e_{2,d^{\prime}} , e_{3,d^{\prime}} \big\} \big\}    \neq        \big\{ e_{1,d^{\prime}} , e_{3,d^{\prime}} \big\}  ,  \big\{ e_{2,d^{\prime}} , e_{3,d^{\prime}}    \big\}    \bigg\}          , 
\\ \\ 
 \textbf{1}_3     \equiv \textbf{1}       \bigg\{      \underset{1 \leq d^{\prime} \leq d }{\mathrm{min}} \big\{  \big\{ e_{1,d^{\prime}}  , e_{2,d^{\prime}} \big\} ,   \big\{         e_{1,d^{\prime}} , e_{3,d^{\prime}} \big\} , \big\{ e_{2,d^{\prime}} , e_{3,d^{\prime}} \big\} \big\}    \neq        \big\{ e_{1,d^{\prime}} , e_{2,d^{\prime}} \big\}  ,  \big\{ e_{1,d^{\prime}} , e_{3,d^{\prime}}    \big\}    \bigg\}          , 
\end{align*}

\noindent with respect to the diamond norm, $d_{\textit{Diamond}} \big\{ \cdot , \cdot \big\} $. Equipped with such quantities, we introduce the following statements for the Main Results below.

\subsubsection{Statement of main results}

\noindent For the main results stated below, we introduce the following additional game-theoretic objects, for establishing a connection between error bounds for the Odd-Cycle game and the number of edges removed from cycles.

\bigskip

\noindent \textbf{Definition} \textit{18} (\textit{the referee's probability distribution in the Odd-Cycle game}). Introduce,

\begin{align*}
    \pi_{\mathrm{Odd-Cycle}} \equiv \underset{i \in V ( \textbf{T}^2 )}{\bigcup} \big\{ i : \textit{the referee distributes i to Alice, or to Bob, for coloring in} \\ \textit{the Odd-Cycle game} \big\} ,
\end{align*}

\noindent corresponding to the referee's probability distribution.

\bigskip

\noindent \textbf{Definition} \textit{19} (\textit{the referee's scoring predicate over Alice and Bob's strategies in the Odd-Cycle game}). The referee's scoring predicate for the Odd-Cycle game, $V_{\mathrm{Odd-Cycle}}$, equals,

{\small \begin{align*}
          \mathcal{S}_A \big( x \big) \oplus \mathcal{S}_B \big( \widetilde{x+t} \big) \equiv  \big[ \mathcal{S}_A \big( x \big) +  \mathcal{S}_B \big( \widetilde{x+t} \big) \big] \text{ } \mathrm{mod}  2 =              \left\{\!\begin{array}{ll@{}l} t \Longleftrightarrow  \textit{Alice and Bob win the Odd-Cycle} \\ \textit{ game,} \\ 0 \text{ } \textit{otherwise,} \end{array}\right. 
                      ,
\end{align*} }

\noindent corresponding to the referee's scoring predicate over $\mathcal{T} \equiv \big( \mathcal{S}_A , \mathcal{S}_B \big)$, for,

\begin{align*}
   \mathcal{S}^d_A , \mathcal{S}^d_B : \big[ n \big]^d \longrightarrow \big[ 2 \big]^d , \\  \mathcal{S}_A , \mathcal{S}_B : \big[ n \big] \longrightarrow \big[ 2 \big] , 
\end{align*}

\noindent where,

\begin{align*}
  \mathcal{S}^d_A \big( x \big) = \mathcal{S}^d_B \big( x \big) = x \text{ } \mathrm{mod} 2  .
\end{align*}

\bigskip

\noindent \textbf{Definition} \textit{20} (\textbf{Definition} \textit{4.2}, [1], \textit{edges removed over the two-dimensional torus blocking topologically nontrivial cycles}). Denote $\mathcal{E}$ as the collection of edges removed over $\textbf{T}^2$ which block all topologically nontrivial cycles, ie the edges $e \in E \big( \textbf{T}^2 \big)$ for which a topologically trivial cycle would exist.

\bigskip

\noindent \textbf{Definition} \textit{21} (\textit{quantification of the number of tensors for Alice and Bob in error inequalities for the Odd-Cycle game}). Fix,

\begin{align*}
 i \neq j \in V \big( \textbf{T}^2 \big)    ,
\end{align*}

\noindent from which we introduce the tensors,

\begin{align*}
  A_i \equiv \textit{Tensor for Alice's response to color vertex i over } \textbf{T}^2 \textit{ drawn uniformly at random } \\ \textit{from the referee's probability distribution} , \\ \\  B_j \equiv  \textit{Tensor for Bob's response to color the neighboring vertex j, to i, over } \textbf{T}^2 \textit{ draw uni-} \\ \textit{  formly at random from the referee's probability distribution}  .
\end{align*}

\noindent From tensors corresponding to Alice and Bob's answers that are scored against $V$, it is possible to consider the ratio,

\begin{align*}
    \frac{\mathrm{sup} \big\{  \omega_q \big( \mathrm{Odd-Cycle} \big)  \big|_{\textit{tensors in } \mathscr{T}_{\textit{contraction}}}     -    \omega_q \big( \mathrm{Odd-Cycle} \big) \big|_{\textit{tensors in } \mathscr{T}^{-1}_{\textit{contraction}}}    \big\} }{\omega_c \big( \mathrm{Odd-Cycle} \big)               }       ,
\end{align*}

\noindent of Quantum, and Classical, winning probabilities, where,

\begin{align*}
 \omega_q \big( \mathrm{Odd-Cycle} \big)  \equiv \textit{Winning probability for the Odd-Cycle game where Alice and Bob make} \\ \textit{ use of Quantum strategies}   , \\ \\ \omega_c \big( \mathrm{Odd-Cycle} \big) \equiv \textit{Winning probability for the Odd-Cycle game where Alice and Bob make} \\ \textit{ use of Classical strategies}  . 
\end{align*}

\noindent \textbf{Definition} \textit{22} (\textit{tensor contraction mapping}). Denote,

\begin{align*}
    \ket{\theta_{+}} \equiv \frac{1}{\sqrt{2}} \bigg[ \ket{0} \pm \mathrm{exp} \big( i \theta \big) \ket{1} \bigg]    ,
\end{align*}

\noindent where $0 \leq \theta \leq 2 \pi$. The mapping,

\begin{align*}
  \mathscr{T} :  E \big( \textbf{T}^2 \big)        \longrightarrow   \underset{\textit{vertices } \in V ( \textbf{T}^2)}{\bigotimes}       \big\{  \textit{Tensor representations of Alice and Bob's answers} \big\} \\  \equiv \underset{\textit{vertices } \in V ( \textbf{T}^2)}{\bigotimes}  \big\{  T_1 \big\}  ,
\end{align*}

\noindent between the edge set of the two-dimensional torus, $\textbf{T}^2$, and the tensor product of Alice and Bob's responses in the Odd-Cycle game induces the \textit{contraction mapping},

{\small \begin{align*}
  \mathscr{T}_{\textit{contraction}} :  \mathcal{E}        \longrightarrow \underset{\textit{vertices } \in V ( \textbf{T}^2)}{\bigotimes} \bigg\{ \bigg\{  \textit{Tensor representations of Alice and Bob's answers} \bigg\}      \backslash  \bigg\{   \textit{Tensor} \\ \textit{representations of Alice and Bob's answers for which they are  unable} \\ \textit{  to compute the angles } \alpha_x \equiv   \frac{\pi x \big( n- 1 \big)}{n} - \frac{\pi}{2 n} , \beta_y \equiv -       \frac{\pi y \big( n- 1 \big)}{n}         \textit{ in the} \\ \textit{ two-outcome projective measurement basis }  \big\{ \ket{\theta_{+}} \bra{\theta_{+}} ,  \ket{\theta_{-}} \bra{\theta_{-}} \big\}   \bigg\}  \bigg\}  \\ \equiv  \underset{\textit{vertices } \in V ( \textbf{T}^2)}{\bigotimes}     \big\{  T_1 \backslash T_2   \big\}         ,
\end{align*}}

\noindent of tensors.

\bigskip

\noindent In the below results, we demonstrate how the ratio,

{\small \begin{align*}
     \frac{ \bigg|\underset{\textit{vertices } \in V ( \textbf{T}^2)}{\bigotimes}        \big\{  T_1  \big\} \bigg| }{\bigg| \underset{\textit{vertices } \in V ( \textbf{T}^2)}{\bigotimes}    \big\{  T_1 \backslash T_2   \big\}  \bigg|  } \equiv    \frac{ \bigg|\underset{\textit{vertices } \in V ( \textbf{T}^2)}{\bigotimes}        \big\{  T_1  \big\} \bigg| }{\bigg| \underset{\textit{vertices } \in V ( \textbf{T}^2)}{\bigotimes}    \big\{  T_1  \big\} \backslash \underset{\textit{vertices } \in V ( \textbf{T}^2)}{\bigotimes} \big\{   T_2 \big\}   \bigg|  }  \\ \equiv   \frac{ \bigg|\underset{\textit{vertices } \in V ( \textbf{T}^2)}{\bigotimes}        \big\{  T_1  \big\} \bigg| }{\bigg| \underset{\textit{vertices } \in V ( \textbf{T}^2)}{\bigotimes}    \big\{  T_1  \big\} \bigg| \backslash  \bigg| \underset{\textit{vertices } \in V ( \textbf{T}^2)}{\bigotimes} \big\{   T_2 \big\}   \bigg|  }     ,
\end{align*}}

\noindent of tensor product spaces in the images of $\mathscr{T}$ and $\mathscr{T}_{\textit{contraction}}$, respectively, correspond to the following three possibilities:wing three possibilities:

\begin{itemize}
\item[$\bullet$] \textit{(1): The ratio of the image of $\mathscr{T}$, with that of $\mathscr{T}_{\textit{contraction}} $, is approximately equal to $1$}. The first possibility would correspond to relatively few \textit{new} tensors belonging to the image of $\mathscr{T}_{\textit{contraction}}$, in comparison to that of $\mathscr{T}$, after contraction.
\item[$\bullet$] \textit{(2): The ratio of the image of $\mathscr{T}$, with that of $\mathscr{T}_{\textit{contraction}} $, is approximately is much larger than $1$}. The second possibility would correspond to the number of tensors contracted in the image $\mathscr{T}_{\textit{contracted}} \big( {E} \big)$, for some $E \in \mathcal{E}$. 
\item[$\bullet$] \textit{(3): The ratio of the image of $\mathscr{T}$, with that of $\mathscr{T}_{\textit{contraction}}$, incorporates behaviors of both (1) and (2) above}. The third possibility would correspond to the existence of a \textit{minimum} number of unmarked connected components, as dictated through \textbf{Definition} \textit{14} of the giant, given below.

\item[$\bullet$] \textit{(4): The ratio of the image of $\mathscr{T}$, with that of $\mathscr{T}_{\textit{contraction}}$, is approximately $0$}. The fourth possibility is never expected to hold, as the number of tensors contained in the image under $\mathscr{T}$ is always expected to be larger than the number of tensors contained in the image of $\mathscr{T}_{\textit{contraction}}$.

\end{itemize}

\bigskip

\noindent \textbf{Definition} \textit{23} (\textit{the marked giant connected component, relative to tubes over the two-dimensional torus}). Introduce,

{\small \begin{align*}
  \mathcal{G} \equiv \underset{\textit{connected components}}{\bigcup}  \text{ } \text{ }  \overset{\cdot}{\underset{\textit{tube:tube} \cap \textbf{T}^2 \neq \emptyset}{\bigcup}}   \bigg\{ \forall \textit{connected components, } \exists \textit{tube} :  \textbf{P} \bigg[ \frac{\big| V \big( \textit{connected components} \big)\big| }{\big| V \big( \textit{tube} \big) \big|}    \\    \approx \frac{95}{100} \bigg] = 1     \bigg\}        , \\ 
\end{align*}}

\noindent corresponding to the \textit{marked} connected component of the giant, for the number of vertices contained in the connected components, $\big| V \big( \textit{connected components} \big)\big|$, relative to the number of vertices in a tube, $\big| V \big( \textit{tube} \big) \big|$, namely,

{\small \begin{align*}
  \big| V \big( \textit{connected components} \big)\big| \equiv \# \big\{  v \in V \big( \textbf{T}^2 \big) : v \textit{ has nonempty intersection with the vertex be-} \\ \textit{longing to a marked connected component contained within a tube over the torus}     \big\}              , \\ \\  \big| V \big( \textit{tube} \big) \big| \equiv   \# \big\{          v \in V \big( \textbf{T}^2 \big) : v \textit{ has nonempty intersection with a vertex belonging to the interior,} \\ \textit{ and also on the boundary, of a tube}  \big\}       . 
\end{align*} }

\bigskip

\noindent \textbf{Definition} \textit{24} (\textit{properties of the marked giant}). $\mathcal{G}$, as defined in the previous item above, satisfies:

\begin{itemize}
    \item[$\bullet$] \textit{Property 1}: The giant has cardinality that contains at most 95 percent of the unmarked vertices in a tube of length $L>0$. 

    \item[$\bullet$] \textit{Property 2}: Suppose that a collection of deleted edges in two sets over the torus contains at most $3 \times 10^{-6}$ edges. Then absolute value distance between the two sets in a tube is equal to two. 

    \item[$\bullet$] \textit{Property 3}: Any collection of marked connected components,

    \begin{align*}
  \underset{\textit{connected component}}{\bigcup}   \big\{ \textit{connected component} : \textit{connected component has a nonempty} \\ \textit{ intersection with an marked vertex over the tube} \big\}        , 
    \end{align*}

\noindent has nonempty intersection with $\mathcal{G}$.

\end{itemize}

\bigskip

\noindent \textbf{Theorem} \textit{1} (\textit{under the assumption that the number of marked vertices in the giant satisfies the above properties, the probability of sampling a tensor contraction mapping Property 3 provided following Definition 13, occurs whp}). With high probability, the event,

{\tiny \begin{align*}
    \bigg\{  \forall \mathcal{G}  , \exists \bigg\{  \big\{ \mathscr{T}_{\textit{contraction}} \big\}  ,   \big\{  1 > \epsilon_1  > 0  \big\} \bigg\}   :  \epsilon_1   <  \frac{\mathrm{sup} \bigg\{  \omega_q \big( \mathrm{Odd-Cycle} \big)  \bigg|_{\textit{tensors in } \mathscr{T}_{\textit{contraction}}}     -    \omega_q \big( \mathrm{Odd-Cycle} \big) \bigg|_{\textit{tensors in } \mathscr{T}^{-1}_{\textit{contraction}}}    \bigg\} }{\omega_c \big( \mathrm{Odd-Cycle} \big)               }       \\      < \frac{1}{\epsilon_1}    \bigg\}     ,
\end{align*}}

\noindent corresponding to $\mathcal{E}_1$ has a probability which can be expressed with,

\begin{align*}
  \textbf{P}_G     \big[   \mathcal{E}_1      \big]              \approx  1 , 
\end{align*}

\noindent where,

{\small \begin{align*}
 \omega_q \big( \mathrm{Odd-Cycle} \big)  \big|_{\textit{tensors in } \mathscr{T}_{\textit{contraction}}}             \equiv             \textit{The probability that Alice and Bob win the Odd-Cycle game} \\ \textit{ with two Quantum strategies, given the constraint that they only make use of tensors} \\ \textit{under the image of }   \mathscr{T}_{\textit{contraction}}   , \\ \\  \omega_q \big( \mathrm{Odd-Cycle} \big)  \big|_{\textit{tensors in } \mathscr{T}^{-1}_{\textit{contraction}}}             \equiv             \textit{The probability that Alice and Bob win the Odd-Cycle game} \\ \textit{ with two Quantum strategies, given the constraint that they only make use of tensors} \\ \textit{under the preimage of }   \mathscr{T}_{\textit{contraction}}           .
\end{align*} }

\bigskip

\noindent In the next result below, to relate the ordinary parallel repetition operation to $\mathscr{T}_{\textit{contraction}}$, as a matter of shorthand, denote,

{\small \begin{align*}
      \omega_q \big( \mathrm{Odd-Cycle}^2 \big)              \big|_{\textit{contraction}} \equiv  \omega_q \big( \mathrm{Odd-Cycle} \big)^{\otimes 2}              \big|_{\textit{contraction}}  , \end{align*}
      
   \begin{align*} \mathrm{sup} \big\{ \omega_q \big( \mathrm{Odd-Cycle}^2 \big) \big\}   \equiv     \mathrm{sup} \big\{ \omega_q \big( \mathrm{Odd} \mathrm{-Cycle} \big)^{\otimes 2} \big\}        . 
\end{align*}}

\noindent \textbf{Theorem} \textit{2} (\textit{under the assumption that the number of marked vertices in the giant satisfies the above properties, under one round of ordinary parallel repetition the probability of sampling a tensor contraction mapping Property 3 provided following Definition 13, occurs whp}). With high probability, the event,

{\tiny \begin{align*}
      \bigg\{   \forall \mathcal{G}  , \exists \bigg\{ \big\{ \mathscr{T}_{\textit{contraction}} \big\} , \big\{   1 >  \epsilon_2 \neq \epsilon_1 > 0 \big\} \bigg\}    : \epsilon_2  <    \frac{\mathrm{sup} \bigg\{  \omega_q \big( \mathrm{Odd-Cycle} \big)^{\otimes 2 }  \bigg|_{\textit{tensors in } \mathscr{T}_{\textit{contraction}}}     -    \omega_q \big( \mathrm{Odd-Cycle} \big)^{\otimes 2} \bigg|_{\textit{tensors in } \mathscr{T}^{-1}_{\textit{contraction}}}    \bigg\} }{\omega_c \big( \mathrm{Odd-Cycle} \big)^{\otimes 2 }               }        \\    <    \frac{1}{\epsilon_2 }        \bigg\}      ,
\end{align*}}

\noindent corresponding to $\mathcal{E}_2$ which can be expressed with,

\begin{align*}
  \textbf{P}_G      \big[ \mathcal{E}_2        \big]                \approx  1 , 
\end{align*}

\noindent where,

\begin{align*}
 \omega_q \big( \mathrm{Odd-Cycle} \big)^{\otimes 2}    \big|_{\textit{tensors in } \mathscr{T}_{\textit{contraction}}}             \equiv               \textit{The probability that Alice and Bob win the Odd-Cycle} \\ \textit{ game with two Quantum strategies, given the constraint that they only make use} \\ \textit{ of tensors under the image of }   \mathscr{T}_{\textit{contraction}}, \textit{ under one application of ordinary parallel} \\ \textit{repetition}     , \\ \\  \omega_q \big( \mathrm{Odd-Cycle} \big)^{\otimes 2}  \big|_{\textit{tensors in } \mathscr{T}^{-1}_{\textit{contraction}}}             \equiv             \textit{The  probability that Alice and Bob win the Odd-Cycle} \\ \textit{ game with two Quantum strategies, given the constraint that they only make use} \\ \textit{ of tensors under the preimage of }   \mathscr{T}_{\textit{contraction}}, \textit{ under one application of ordinary parallel} \\ \textit{repetition}     .
\end{align*}

\bigskip

\noindent \textbf{Proposition} (\textit{expressing the above probabilities with probabilities relating to surface area problems of foams}. The probability introduced above in \textbf{Theorem} \textit{2}, and hence the probability introduced in \textbf{Theorem} \textit{1}, can be related to the probability,

{\small \begin{align*}
     \textbf{P}_G     \bigg[  \forall \mathcal{G} , \exists  \bigg\{ \big\{ \textit{foam} \big\} , \big\{ \textit{tube} \big\}  , \big\{ \textit{section} \big\} , \big\{ m > 0 \big\} , \big\{ d \equiv 2 \big\}  \bigg\} : \big\{ \textit{(1)}  \big\} , \big\{       \textit{(2)}                \big\}     ,        \big\{      \textit{(3)}    \big\}     ,     \big\{     \textit{(4)}    \big\}  \text{ } \textit{simul-} \\ \textit{taneously occur}    \bigg]               ,
\end{align*}    }

\noindent through the ratio,

{\small \begin{align*}
 \textbf{P}_G      \big[ \mathcal{E}_2        \big] \textbf{P}_G     \bigg[  \forall \mathcal{G} , \exists  \bigg\{ \big\{ \textit{foam} \big\} , \big\{ \textit{tube} \big\}  , \big\{ \textit{section} \big\} , \big\{ m > 0 \big\} , \big\{ d \equiv 2 \big\}  \bigg\} : \big\{ \textit{(1)}  \big\} , \big\{       \textit{(2)}                \big\}     ,        \big\{      \textit{(3)}    \big\}     ,     \big\{     \textit{(4)}    \big\}  \text{ } \textit{simul-} \\ \textit{taneously occur}    \bigg]^{-1}              ,
\end{align*}    }

\noindent up to a constant which equals the following product of factors,

\begin{align*}
   \mathscr{F}_1     \mathscr{F}_2          \mathscr{F}^{-1}_3 \mathscr{F}_4  \mathscr{F}_5 \mathscr{F}^{-1}_6                   ,
\end{align*}

\noindent for,

{\small \begin{align*}
 \mathscr{F}_1 \equiv  \#  \big\{ \textit{tensor} : \textit{tensor is in the image of } \mathscr{T}_{\textit{contraction}} \big\}                            , \end{align*}

 \begin{align*} \mathscr{F}_2 \equiv    \# \big\{ v : v \in  V \big( \textit{tube} \big) \big\} , \end{align*}

 \begin{align*} \mathscr{F}_3 \equiv  \# \big\{ v : v \in V \big( \textbf{T}^2 \big) \big\}            , \end{align*}

 \begin{align*}      \mathscr{F}_4 \equiv \# \big\{ \textit{tensor}: \textit{tensor is contained in the preimage of $\mathscr{T}_{\textit{contraction}}$}   \big\}    , \end{align*}

 \begin{align*}          \mathscr{F}_5    \equiv \# \big\{        v : v \in V \big( \mathcal{G} \big) \big\}          , 
\end{align*}

 \begin{align*}          \mathscr{F}_6    \equiv \# \big\{        v : v \in V \big( \textit{section} \big) \big\}          , 
\end{align*} } 

\noindent corresponding to the number of tensors in the image of $\mathscr{T}_{\textit{contraction}}$, the number of vertices contained within a tube, the number of vertices over $\textbf{T}^2$, the number of tensors contained in the preimage of $\mathscr{T}_{\textit{contraction}}$, the number of vertices of the marked giant, and the number of vertices in a section, respectively, where each event appearing in the union,

\begin{align*}
    \big\{ \textit{(1)}  \big\} \cup  \big\{       \textit{(2)}                \big\}     \cup         \big\{      \textit{(3)}    \big\}     \cup     \big\{     \textit{(4)}    \big\}  , 
\end{align*}

\noindent that is measurable wrt $\textbf{P}_G \big[ \cdot \big]$ is denoted with,

\begin{align*}
   \textit{(1)} \equiv   \textit{foam} \cap \textbf{T}^2 \neq \emptyset , \end{align*}
   
   \begin{align*} \textit{(2)} \equiv  \underset{\textit{each vertex of a foam} \subsetneq V ( \textbf{T}^2 )} {\bigcup}   \big|  \textit{vertex of each foam} \big| \leq   m^{-d} \big[ \# \big\{ v :  v \in \big[  V \big( \textbf{T}^2 \big)                \cap  V \big( \textit{tube} \big) \big] \big\}  \big]    , \end{align*}

   \begin{align*}  \textit{(3)} \equiv \textit{section} \subsetneq \textit{tube}              , \end{align*}
   
   \begin{align*} \textit{(4)} \equiv     d_{\textit{Diamond}}      \bigg\{  \underset{\textit{Surface Area}}{\mathrm{inf}}     \big\{ e_{1,d}  , e_{2,d}  \big\}        \bigg\}     \lesssim n^d   . 
\end{align*}

\bigskip

\noindent \textbf{Theorem} \textit{3} (\textit{the probability that a collection of marked vertices over a section strictly contained within a tube, given the existence of marked connected components of the giant, implies that parallel repetition of the Odd-Cycle game guarantees an up to constant upper bound of order $n^2$ for the surface area of a foam over} $\textbf{T}^2$). The event,

{\tiny \begin{align*}
   \bigg\{    \forall \mathcal{G}  , \exists   \bigg\{   \big\{ \mathscr{T}_{\textit{contraction}} \big\} , \big\{ 1 >  \epsilon_3 \neq \epsilon_2 \neq \epsilon_1  > 0 \big\} \bigg\}  : \frac{1}{\epsilon_3}    <   \frac{\mathrm{sup} \bigg\{ \omega_q \big( \mathrm{Odd-Cycle} \big)^{\otimes 2} - \omega_q \big( \mathrm{Odd-Cycle} \big)^{\otimes 2}              \bigg|_{\textit{contraction}}  \bigg\} }{\omega_q \big( \mathrm{Odd-Cycle} \big)^{\otimes 2}              \bigg|_{\textit{contraction}} }           \\  <   \epsilon_3     \bigg\}  ,
\end{align*}}

\noindent corresponding to $\mathcal{E}_3$ which can be expressed with,

{\small \begin{align*}
    \textbf{P}_G \big[ \mathcal{E}_3 \big]   \textbf{P}_G     \bigg[  \forall \mathcal{G} , \exists  \bigg\{ \big\{ \textit{foam} \big\} , \big\{ \textit{tube} \big\}  , \big\{ \textit{section} \big\} , \big\{ m > 0 \big\} , \big\{ d \equiv 2 \big\}  \bigg\} : \big\{ \textit{(1)}  \big\} , \big\{       \textit{(2)}                \big\}     ,        \big\{      \textit{(3)}    \big\}     ,     \big\{     \textit{(4)}    \big\}  \text{ } \textit{simul-} \\ \textit{taneously occur}    \bigg]^{-1}       \approx 1 .
\end{align*}}

\noindent Intuitively, the above results can be regarded through the following quantitative observations:

\begin{itemize}
\item[$\bullet$] \textit{(1). The probability that the Quantum value of the Odd-Cycle game, compared to the optimal value under the tensor contraction map, is at most of order $\epsilon^{-1}_1$, for strictly positive $\epsilon_1$.} 

\item[$\bullet$] \textit{(2). Under parallel repetition, the above ratio of Quantum optimal values, under one round of ordinary parallel repetition, is at most of order $\epsilon^{-1}_2$, for strictly positive $\epsilon_2$.}

\item[$\bullet$] \textit{(3). Relating the above two probabilities to the minimizing surface area of a foam over the torus}.

\item[$\bullet$] \textit{(4). Relating the probabilities that the Quantum value of the Odd-Cycle game, under one ordinary parallel repetition operation, is approximately of the same order as the probability that there exists a foam over the torus with minimizing surface area}.

\end{itemize}

\noindent Before providing the arguments for each of the main results stated above in \textit{3}, in \textit{2.1} and \textit{2.2} in the next section we formally introduce several topological, and geometric, objects.

\section{Game-theoretic objects}

\subsection{Topology of the Odd-Cycle game, consistent pearls, winning strategies}

\noindent We define several objects that have previously been used to topologically characterize the Odd-Cycle game appearing in [1].

\bigskip

\noindent \textbf{Definition} \textit{25} (\textit{value of the CHSH game after three applications of the parallel repetition operation}). Under three rounds of the ordinary parallel repetition operation, the Quantum value, $\omega_q$, of the CHSH game equals,

\begin{align*}
    \omega_q  \big( \textit{CHSH} \big)^{\otimes 3} =  \omega_q  \big( \textit{CHSH}^3 \big) =  \frac{31}{64}  . 
\end{align*}

\noindent \textbf{Definition} \textit{26} (\textit{consistent regions}). A region, $R \subseteq \big[ n \big]^d$, is said to be \textit{consistent}, with respect to Alice's strategy $\mathcal{S}_A$, if,

\begin{align*}
 \mathcal{S}_A \big( x \big) -    \mathcal{S}_A \big( x^{\prime} \big) = \big( x - \widetilde{x^{\prime}} \big) \text{ } \mathrm{mod} 2  . 
\end{align*}

\noindent \textbf{Definition} \textit{27} (\textit{the value of the Odd-Cycle game as a summation over consistent regions}). Given consistent regions defined in the previous \textbf{Definition} above, one can express the optimal value as,

\begin{align*}
 \omega_q \big( \mathrm{Odd-Cycle} \big) =    \frac{1}{n^d 2^d} \underset{y \in [ n ]^d}{\sum}   \big| R_y \big|       . 
\end{align*}

\bigskip

\noindent \textbf{Definition} \textit{28} (\textit{pearls that are ordinary, and pearls that are consistent, over the two-dimensional torus}). An \textit{ordinary} pearl, ie a collection,

\begin{align*}
   \rho \equiv \big\{ R_y \big| y \in \big[ n \big]^d \big\}  ,
\end{align*}

\noindent which satisfies,

\begin{align*}
  R_y \subseteq Q_y , \forall y \in \big[ n \big]^d   . 
\end{align*}

\noindent With respect to Alice's strategy, $\mathcal{S}_A$, a \textit{consistent} pearl, besides satisfying the above construction, also satisfies,

\begin{align*}
    \textit{Consistent pearls} \equiv        \underset{y}{\bigcup} \big\{ \textit{pearls: regions } R_y \textit{ strictly contained within the pearl are consistent} \big\}                . 
\end{align*}

\noindent \textbf{Definition} \textit{29} (\textit{the set of odd blockers}). Introduce,

\begin{align*}
\mathscr{B} \equiv  \big\{  \textit{Blocker : Blocker has odd length, and all vertices contained within the Blocker have} \\ \textit{ vanishing homotopy}     \big\}    , 
\end{align*}

\noindent corresponding to the set of all odd blockers.

\bigskip

\noindent \textbf{Theorem} (\textbf{Theorem} \textit{3}, [1], \textit{the winning probability of the Odd-Cycle game admits an integral representation over the set of odd blockers}). Fix strictly positive $n$ for which,

\begin{align*}
  \big[ n \times \mathscr{B} \big]  \cap \big[ n \big]^d = \emptyset ,
\end{align*}

\noindent $\epsilon, n_{\epsilon}$ sufficiently small. For $n \geq n_{\epsilon}$, one has that,

\begin{align*}
     \omega_q \big( \textit{Pearls of the }  n   \text{ } \textit{length Odd Blocker} \text{ } \mathscr{B} \big)  \geq   1 - \frac{1 + \epsilon}{n} \underset{\mathscr{B}}{\int \int}  \big| \mathrm{d} \mathscr{B} \big|_{\textit{Diamond}}  .
\end{align*}

\noindent The above winning probability associated with the pearl of an odd blocker can be used for constructing closely related objects that satisfying the following, [18]:

\begin{itemize}
  \item[$\bullet$] \textit{Definition of a measure over the winning prob abilities associated with Alice's strategy}. Define the measure,

        \begin{align*}
           \lambda \big( S \big) \equiv \underset{n \longrightarrow + \infty}{\mathrm{lim}}  n \bigg[ 1 - \frac{1}{2^d n^d} \underset{y \in [ n]^d }{\sum}  \big| R_y \big|           \bigg]        .
        \end{align*}

   \item[$\bullet$] \textit{Additivity}. One has that a decomposition of the below form, corresponding to the measure over $S$,

    \begin{align*}
       \lambda \big( S \big) = \underset{1 \leq i \leq m}{\sum} \lambda \big( S_i \big)  \equiv  \underset{1 \leq i \leq m}{\sum} \big\{  \lambda \textit{ over each } S_i \big\}  ,
    \end{align*}

   \noindent holds, corresponding to \textit{additivity} of the measures over each $S_i$, for all $i$.

   \item[$\bullet$] \textit{Decomposition of the measure over each } $S_i$ \textit{in terms of the diamond norm}). Write $S = \big[ s_1 , \cdots , s_d \big]$, and recall the definition of random variables $\chi$ in \textbf{Definition} \textit{12}. Introduce the decomposition,

\begin{align*}
     \lambda \big( S \big) \equiv \frac{1}{2} E \bigg[ \bigg|  \underset{1 \leq i \leq d}{\sum}    s_i \chi_i         \bigg|     \bigg]    \overset{(\textit{Triangle inequality})}{\leq}      \frac{1}{2} E \bigg[   \underset{1 \leq i \leq d}{\sum}  \big|   s_i \chi_i         \big|     \bigg]             , 
\end{align*}

   \noindent from the diamond norm, where,

   \begin{align*}
 E \big[ \cdot \big] \equiv \textit{Expectation with respect to the probability measure } \textbf{P}_G \big[ \cdot \big]   .
   \end{align*}

\end{itemize}

\noindent Lastly, one can characterize cycles with odd length, through the following property.

\bigskip

\noindent \textbf{Definition} \textit{30} (\textit{topological non-triviality, and topological oddity}). For $n$ odd and a cycle $C \equiv x_0, \cdots , x_k \in \big[ n \big]^d$, with $x_k \equiv x_0$,

\begin{align*}
            \textit{The cycle is topologically non-trivial}  \Longleftrightarrow            \underset{1 \leq i \leq k-1}{\sum} \big[ \widetilde{x_{i+1} - x_i} \big]  \neq 0            , \\ \\         \textit{The cycle is topologically odd}  \Longleftrightarrow                 \underset{1 \leq i \leq k-1}{\sum} \big[ \widetilde{x_{i+1} - x_i} \big] \neq 0  \text{ } \mathrm{mod} 2               . 
\end{align*}

\subsection{Extracting topological characteristics of pearls, odd blockers, and the winning probability under $\mathcal{S}_A$}

\noindent Denote,

\begin{align*}
    \textbf{P}_G \big[ \cdot \big]    ,
\end{align*}

\noindent corresponding to the probability measure whose support is over graphs $G$ whose edges block odd cycles. Equipped with the objects provided in the previous subsection, probabilistically we reformulate arguments provided in [18] for  $v \big( G^2_n \big)$ through the following quantities:

\begin{itemize}
    \item[$\bullet$] \textit{Initialization}. Define an empty cycle $\mathscr{C}_0 \equiv {\emptyset}$, along with a \textit{torical} graph, $G$ (namely, a graph whose edges block all odd cycles), and an orientation of $\Gamma$.

      \item[$\bullet$] \textit{Increasing the number of points in the cycle}. Over $G$, the number of points in each cycle, given the initialization provided in the previous item above, can be obtained by constructing a sequence of cycles,

      \begin{align*}
        \big\{ \mathscr{C}_i \big\}_{1 \leq i \leq N}  ,
      \end{align*}

      \noindent for some strictly positive $N$, whose cardinalities satisfy,

      \begin{align*}
       N \equiv \big| \mathscr{C}_N \big| > N - 1 \equiv \big| \mathscr{C}_{N-1} \big| \equiv \big| \mathscr{C}_N \big| - 1 > \cdots > 0 \equiv \big| \mathscr{C}_{\emptyset} \big|  \equiv   \big| \mathscr{C}_1 \big| - 1  .
     \end{align*}

        \item[$\bullet$] \textit{Increasing the number of points in a cycle within consistent regions}. Given previous definitions of consistent regions, in addition to consistent pearls, one is able to increase the length of the cycle iff, with respect to $\mathcal{S}_A$,

    \begin{align*}
     \textbf{P}_G \bigg[             \forall \textit{consistent } Q_y, \exists                  x^* \in Q^C_y \cap G : L_{\infty} \big\{   x^*   ,  Q^C_y   \big\}   \equiv \underset{z \in Q^C_y \cap G}{\mathrm{min}}                  L_{\infty} \big\{  z    ,   Q_y  \big\}       \bigg]   > 0    ,
    \end{align*}

\noindent for the $L$-$\infty$ norm, $L_{\infty} \big\{ \cdot , \cdot \big\} $.

  \item[$\bullet$] \textit{Determining whether the newly added point to the cycle is consistent}. Necessarily,

  {\small\begin{align*}
   \textbf{P}_G \bigg[  \forall 1 \leq i \leq N, \exists x^* \in Q^C_y \cap G : \bigg\{       x_0 , \cdots  , x_i , x^* \textit{ is a cycle} \text{ }  C_i \cup \big\{ x^* \big\}   \bigg\} , \bigg\{  \big\{   C_i  \cup \big\{ x^*  \big\}      \\  \subseteq Q_y        \bigg\}     \bigg]    > 0  , 
  \end{align*}}

  \noindent corresponding to the observation that, with strictly positive probability, the newly added point in the cycle is consistent with all of the previously added points of the cycle.

    \item[$\bullet$] \textit{Topological evenness}. The aforementioned procedure for adding new points to the cycle implies,

    \begin{align*}
       \textbf{P}_G \big[ \textit{The cycle, with the point } x^* \textit{added, is topologically even} \big]  = 1 .
    \end{align*}

    \item[$\bullet$] \textit{Vanishing homotopy}. Besides the topological property of the cycle asserted in the previous item, one also has that,

    \begin{align*}
       \textbf{P}_G \big[    \textit{The cycle } \Gamma,  \text{ }  \textit{with a prescribed clockwise, or counterclockwise, orientation around} \\ \textit{ the two-dimensional torus, has homotopy zero}             \big] = 1 .
    \end{align*}

    \item[$\bullet$] \textit{The isoperimetric inequality between} $n$ \textit{and the winning probability under Alice's strategy}. Concluding, one has that,

    \begin{align*}
          \textbf{P}_G \bigg[  \forall \mathcal{S}_A , \exists n > 0 :       1.5n \leq 2 n^2 \big[ 1 - v \big( \mathcal{S}_A \big) \big]          \bigg] = 1 . 
    \end{align*}

\end{itemize}

\noindent For the purposes of the upcoming arguments, given the fact that $d \equiv 2$, we do not make use of the above isoperimetric inequality which is used for large $d$. It could very well be of interest in the future to determine how generalizations of such inequalities could be put to use.

\section{Arguments for the Main results}

\subsection{Theorem $\textit{1}$}

\noindent \textit{Proof of Theorem 1}. To demonstrate that the event provided in the statement of the result occurs whp, it suffices to construct the sequence,

{\tiny \begin{align*}
      \bigg\{            \forall \mathcal{G}        ,   \exists \bigg\{  \big\{ \mathscr{T}_{\textit{contraction}} \big\}  ,   \big\{  1 > \epsilon^{\prime}_1  > 0  \big\} \bigg\}   :  \epsilon^{\prime}_1   <  \frac{\mathrm{sup} \bigg\{  \omega_q \big( \mathrm{Odd-Cycle} \big)  \bigg|_{\textit{tensors in } \mathscr{T}_{\textit{contraction}}}     -    \omega_q \big( \mathrm{Odd-Cycle} \big) \bigg|_{\textit{tensors in } \mathscr{T}^{-1}_{\textit{contraction}}}    \bigg\} }{\omega_c \big( \mathrm{Odd-Cycle} \big)               }       \\      < \frac{1}{\epsilon^{\prime}_1}          \bigg\}_{\epsilon^{\prime}_1 \leq \epsilon_1}                   ,
\end{align*}}

\noindent of events corresponding to $\mathcal{E}^{\prime}_1 \equiv \underset{\epsilon^{\prime}_1 \in \textbf{R} : \epsilon^{\prime}_1 \leq \epsilon_1}{\bigcup} \mathcal{E}^{\prime}_1 \big( \epsilon^{\prime}_1 \big)$, and hence the sequence,

\begin{align*}
    \big\{ \textbf{P}_G \big[  \mathcal{E}^{\prime}_1  \big] \big\}_{\epsilon^{\prime}_1 \leq \epsilon_1}         ,
\end{align*}

\noindent of probabilities, each of which has strictly positive mass with respect to $\textbf{P}_G \big[ \cdot \big]$. That is, one expects the existence of a threshold, $\Theta$, for which,

{\small 
\[
\left\{\!\begin{array}{ll@{}>{{}}l}               \frac{  \big\{ \textbf{P}_G \big[  \mathcal{E}^{\prime}_1  \big] \big\}_{\epsilon^{\prime}_1 \leq \Theta_1} }{\textbf{P}_G \big[ \mathcal{E}_1  \big]}   =        \bigg\{         \frac{   \textbf{P}_G \big[  \mathcal{E}^{\prime}_1  \big]  }{\textbf{P}_G \big[ \mathcal{E}_1  \big]}    \bigg\}_{\epsilon^{\prime}_1 \leq \Theta_1}        \approx                       1         \Longleftrightarrow  \Theta_1 < \Theta   ,    \\ \frac{  \big\{ \textbf{P}_G \big[  \mathcal{E}^{\prime}_1  \big] \big\}_{\epsilon^{\prime}_1 \leq \Theta_1} }{\textbf{P}_G \big[ \mathcal{E}_1  \big]}       =   \bigg\{ \frac{  \textbf{P}_G \big[  \mathcal{E}^{\prime}_1  \big] }{\textbf{P}_G \big[ \mathcal{E}_1  \big]}    \bigg\}_{\epsilon^{\prime}_1 \leq \Theta_1}     \approx   1 + \mathrm{o} \big(    \Theta_1 \textbf{1}_{ \{ \Theta_1 : \Theta_1 > \Theta \} }     \big) \approx 1     \Longleftrightarrow  \Theta_1 > \Theta , \mathrm{o} \big(    \Theta_1 \textbf{1}_{ \{ \Theta_1 : \Theta_1 > \Theta \} }     \big)  \longrightarrow 0  \\ \textit{as } n \longrightarrow + \infty  . 
\end{array}\right.
\] }

\noindent To lighten the notation appearing in the forthcoming computations, denote,

\begin{align*}
    \mathscr{R} \equiv \frac{\mathrm{sup} \big\{  \omega_q \big( \mathrm{Odd-Cycle} \big)  \big|_{\textit{tensors in } \mathscr{T}_{\textit{contraction}}}     -    \omega_q \big( \mathrm{Odd-Cycle} \big) \big|_{\textit{tensors in } \mathscr{T}^{-1}_{\textit{contraction}}}    \big\} }{\omega_c \big( \mathrm{Odd-Cycle} \big)               }     . 
\end{align*}

\noindent Furthermore, to argue for the existence of a collection of parameters,

\begin{align*}
        \underset{\textit{vertices } n: \text{ } n  \textit{belongs to a topologically nontrivial odd-cycle}}{\bigcup} \big\{   \epsilon_{\mathrm{min}} \equiv  \epsilon_{\mathrm{min}} \big(   V \big( \textbf{T}^2 \big) ,   n   \big)       \big)  : \epsilon_{\mathrm{min}} \leq \epsilon_1 \big\} 
\end{align*}

\noindent so that,

\begin{align*}
    \big\{ \textbf{P}_G \big[  \mathcal{E}^{\prime}_1  \big] \big\}_{\epsilon_{\mathrm{min}} \leq \epsilon_1} \longrightarrow \textbf{P}_G   \big[  \mathcal{E}_1  \big]        ,
\end{align*}

\noindent it suffices to argue, by direct computation, that events of the form,

{\tiny\begin{align*}
        \bigg\{  \forall \mathcal{G}  , \exists \bigg\{  \big\{ \mathscr{T}_{\textit{contraction}} \big\}  ,   \big\{  1 > \epsilon_1  > 0  \big\} \bigg\}   :  \epsilon_1   <  \frac{\mathrm{sup} \bigg\{  \omega_q \big( \mathrm{Odd-Cycle} \big)  \big|_{\textit{tensors in } \mathscr{T}_{\textit{contraction}}}     -    \omega_q \big( \mathrm{Odd-Cycle} \big) \big|_{\textit{tensors in } \mathscr{T}^{-1}_{\textit{contraction}}}    \bigg\} }{\omega_c \big( \mathrm{Odd-Cycle} \big)               }      \\       < \frac{1}{\epsilon_1}    \bigg\}             ,
\end{align*}}

\noindent implies that one can perform the following computation for,

\begin{align*}
   \bigg\{         \frac{   \textbf{P}_G \big[  \mathcal{E}^{\prime}_1 \big( \epsilon^{\prime}_1 \big)   \big]  }{\textbf{P}_G \big[ \mathcal{E}_1  \big]}    \bigg\}_{\epsilon^{\prime}_1 \leq \Theta_1}    \equiv    \bigg\{         \frac{   \textbf{P}_G \big[  \mathcal{E}^{\prime}_1  \big]  }{\textbf{P}_G \big[ \mathcal{E}_1  \big]}    \bigg\}_{\epsilon^{\prime}_1 \leq \Theta_1}           , 
\end{align*}

\noindent can be obtained with the following arguments. First, observe,

{\small \begin{align*}
    \bigg\{       \frac{ \textbf{P}_G  \bigg[  \forall \mathcal{G}  , \exists \bigg\{  \big\{ \mathscr{T}_{\textit{contraction}} \big\}  ,   \big\{  1 > \epsilon^{\prime}_1  > 0  \big\} \bigg\}   :  \epsilon^{\prime}_1   <  \frac{\mathrm{sup} \big\{ \omega_q ( \mathrm{Odd-Cycle} ) \big\} }{\omega_q ( \mathrm{Odd-Cycle} )              \big|_{\textit{contraction}} }            < \frac{1}{\epsilon^{\prime}_1}    \bigg]   }{   \textbf{P}_G   \bigg[  \forall \mathcal{G}  , \exists \bigg\{  \big\{ \mathscr{T}_{\textit{contraction}} \big\}  ,   \big\{  1 > \epsilon_1  > 0  \big\} \bigg\}   :  \epsilon_1   <  \frac{\mathrm{sup} \big\{ \omega_q ( \mathrm{Odd-Cycle} ) \big\} }{\omega_q ( \mathrm{Odd-Cycle} )              \big|_{\textit{contraction}} }            < \frac{1}{\epsilon_1}    \bigg]     }   \bigg\}_{\epsilon^{\prime}_1 \leq \Theta_1}  \end{align*}

    \begin{align*} \approx        \bigg\{    \textbf{P}_G  \bigg[  \forall \mathcal{G}  , \exists \bigg\{  \big\{ \mathscr{T}_{\textit{contraction}} \big\}  ,   \big\{  1 > \frac{\epsilon^{\prime}_1}{\epsilon_1 }  > 0  \big\} \bigg\}   :    \frac{\epsilon^{\prime}_1}{\epsilon_1}      <  \frac{\mathrm{sup} \big\{ \omega_q ( \mathrm{Odd-Cycle} ) \big\} }{\omega_q ( \mathrm{Odd-Cycle} )              \big|_{\textit{contraction}} }        \\     <         \frac{\epsilon_1}{\epsilon^{\prime}_1}    \bigg]  \bigg\}_{\epsilon^{\prime}_1 \leq \Theta_1} . \tag{$\mathcal{P}$} \end{align*}}

\noindent Proceeding, further rearrangement of the above probabilities imply,

   {\small \begin{align*} (\mathcal{P}) <          \textbf{P}_G  \bigg[  \forall \mathcal{G}  , \exists \bigg\{  \big\{ \mathscr{T}_{\textit{contraction}} \big\}  ,   \big\{  1 > \frac{\Theta_1 - 1 }{\Theta_1 }  > 0  \big\} \bigg\}   :    \frac{\Theta_1 - 1 }{\Theta_1}       <  \mathscr{R}        <         \frac{\Theta_1 }{\Theta_1 - 1}    \bigg]                \\ \\        <     \underset{\Theta_1 : 1 > \frac{\Theta_1 - 1 }{\Theta_1 } > 0 }{\mathrm{sup}}  \bigg\{    \textbf{P}_G  \bigg[  \forall \mathcal{G}  , \exists \bigg\{  \big\{ \mathscr{T}_{\textit{contraction}} \big\}  ,   \big\{  1 > \frac{\Theta_1 - 1 }{\Theta_1 }  > 0  \big\} \bigg\}   :    \frac{\Theta_1 - 1 }{\Theta_1}       <  \mathscr{R}       \\   <         \frac{\Theta_1 }{\Theta_1 - 1}    \bigg]    \bigg\} \\        \\   \approx        \underset{\Theta_1 : 1 > \frac{\Theta_1 - 1 }{\Theta_1 } > 0 }{\mathrm{sup}}  \bigg\{    \textbf{P}_G  \bigg[  \forall \mathcal{G}  , \exists \bigg\{  \big\{ \mathscr{T}_{\textit{contraction}} \big\}  ,   \big\{  1 > \frac{\Theta_1 - 1 }{\Theta_1 }  > 0  \big\} \bigg\}   :    \frac{\Theta_1 - 1 }{\Theta_1}      <  \mathscr{R}        \\  \leq          \frac{\Theta_1 }{\Theta_1 - 1}    \bigg]    \bigg\}                \\ \\ =   \underset{\Theta_1 : 1 > \frac{\Theta_1 - 1 }{\Theta_1 } > 0 }{\mathrm{sup}}  \bigg\{    \textbf{P}_G  \bigg[  \forall \mathcal{G}  , \exists \bigg\{  \big\{ \mathscr{T}_{\textit{contraction}} \big\}  ,   \big\{  1 > \frac{\Theta_1 - 1 }{\Theta_1 }  > 0  \big\} \bigg\}   :    \frac{\Theta_1 - 1 }{\Theta_1}       <  \mathscr{R}      \end{align*}

   \begin{align*} <          \frac{\Theta_1 }{\Theta_1 - 1}    \bigg]  +   \textbf{P}_G  \bigg[  \forall \mathcal{G}  , \exists \bigg\{  \big\{ \mathscr{T}_{\textit{contraction}} \big\}  ,   \big\{  1 > \frac{\Theta_1 - 1 }{\Theta_1 }  > 0  \big\} \bigg\}   :         \mathscr{R}        \\   =           \frac{\Theta_1 }{\Theta_1 - 1}    \bigg]        \bigg\}        .  \\  \tag{$\mathcal{P}^{*}$}  \end{align*} }

    \noindent To upper bound the supremum over $\Theta_1$ of the two above probabilities, observe,

{\small \begin{align*}
   (\mathcal{P}^{*})  <    \underset{\Theta_1 : 1 > \frac{\Theta_1 - 1 }{\Theta_1 } > 0 }{\mathrm{sup}}  \bigg\{     \textbf{P}_G  \bigg[  \forall \mathcal{G}  , \exists \bigg\{  \big\{ \mathscr{T}_{\textit{contraction}} \big\}  ,   \big\{  1 > \frac{\Theta_1 - 1 }{\Theta_1 }  > 0  \big\} \bigg\}   :    \frac{\Theta_1 - 1 }{\Theta_1}    \   <  \mathscr{R}        \\  <          \frac{\Theta_1 }{\Theta_1 - 1}    \bigg]     \\      +     \textbf{P}_G  \bigg[  \forall \mathcal{G}   , \exists \bigg\{  \big\{ \mathscr{T}_{\textit{contraction}} \big\}   ,   \big\{  1 > \frac{\Theta_1 - 1 }{\Theta_1 }  > 0  \big\} \bigg\}   :         \mathscr{R}         =           \frac{\Theta_1 }{\Theta_1 - 1}    \bigg]       + \textit{Probability 3}    \bigg\}                     ,
\end{align*}} 

\noindent where,

{\small \begin{align*}
 \textit{Probability 3}     \equiv    \textbf{P}_G  \bigg[  \forall \mathcal{G}   , \exists \bigg\{  \big\{ \mathscr{T}_{\textit{contraction}} \big\}   ,   \big\{  1 > \frac{\Theta_1 - 1 }{\Theta_1 }  > 0  \big\} \bigg\}   :       \mathscr{R}   \textit{is strictly larger} \\ \textit{than } \frac{\Theta_1 - 1}{\Theta_1 } \bigg] 
\end{align*} }

\noindent Hence,

  {\tiny  \begin{align*}  (\mathcal{P}^{*})       < \underset{\Theta_1 : 1 > \frac{\Theta_1 - 1 }{\Theta_1 } > 0 }{\mathrm{sup}}  \bigg\{     \textbf{P}_G  \bigg[  \forall \mathcal{G}  , \exists \bigg\{  \big\{ \mathscr{T}_{\textit{contraction}} \big\}  ,   \big\{  1 > \frac{\Theta_1 - 1 }{\Theta_1 }  > 0  \big\} \bigg\}   :    \frac{\Theta_1 - 1 }{\Theta_1}       <  \mathscr{R}         <          \frac{\Theta_1 }{\Theta_1 - 1}    \bigg] \\   +                      \textbf{P}_G  \bigg[  \forall \mathcal{G}  , \exists \bigg\{  \big\{ \mathscr{T}_{\textit{contraction}} \big\}  ,   \big\{  1 > \frac{\Theta_1 - 1 }{\Theta_1 }   > 0  \big\} \bigg\}    :         \mathscr{R}          =           \frac{\Theta_1 }{\Theta_1 - 1}    \bigg]        \bigg\}            \\  +    \textbf{P}_G  \bigg[  \forall \mathcal{G}  , \exists \bigg\{  \big\{ \mathscr{T}_{\textit{contraction}} \big\}  ,   \big\{  1 > \frac{\Theta_1 - 1 }{\Theta_1 }  > 0  \big\} \bigg\}   :    \frac{\Theta_1 - 1 }{\Theta_1}      <  \mathscr{R}         \bigg]              \bigg\}             \\ \\   \approx      \underset{\Theta_1 : 1 > \frac{\Theta_1 - 1 }{\Theta_1 } > 0 }{\mathrm{sup}}  \bigg\{     \textbf{P}_G  \bigg[  \forall \mathcal{G}  , \exists \bigg\{  \big\{ \mathscr{T}_{\textit{contraction}} \big\}  ,   \big\{  1 > \frac{\Theta_1 - 1 }{\Theta_1 }  > 0  \big\} \bigg\}   :    \frac{\Theta_1 - 1 }{\Theta_1}       <  \mathscr{R}         <          \frac{\Theta_1 }{\Theta_1 - 1}    \bigg]  \bigg\}   \\  +    \underset{\Theta_1 : 1 > \frac{\Theta_1 - 1 }{\Theta_1 } > 0 }{\mathrm{sup}}  \bigg\{                       \textbf{P}_G  \bigg[  \forall \mathcal{G}  , \exists \bigg\{  \big\{ \mathscr{T}_{\textit{contraction}} \big\}  ,   \big\{  1 > \frac{\Theta_1 - 1 }{\Theta_1 }   > 0  \big\} \bigg\}    :         \mathscr{R}          =           \frac{\Theta_1 }{\Theta_1 - 1}    \bigg]        \bigg\}   \end{align*}

   \begin{align*}   +      \underset{\Theta_1 : 1 > \frac{\Theta_1 - 1 }{\Theta_1 } > 0 }{\mathrm{sup}}  \bigg\{                \textbf{P}_G  \bigg[  \forall \mathcal{G}  , \exists \bigg\{  \big\{ \mathscr{T}_{\textit{contraction}} \big\}  ,   \big\{  1 > \frac{\Theta_1 - 1 }{\Theta_1 }  > 0  \big\} \bigg\}   :    \frac{\Theta_1 - 1 }{\Theta_1}      <  \mathscr{R}         \bigg]              \bigg\}  \\ \\ \approx          \underset{\Theta_1 : 1 > \frac{\Theta_1 - 1 }{\Theta_1 } > 0 }{\mathrm{sup}}  \bigg\{     \textbf{P}_G  \bigg[  \forall \mathcal{G}  , \exists \bigg\{  \big\{ \mathscr{T}_{\textit{contraction}} \big\}  ,   \big\{  1 > \frac{\Theta_1 - 1 }{\Theta_1 }  > 0  \big\} \bigg\}   :    \frac{\Theta_1 - 1 }{\Theta_1}       <  \mathscr{R}         <          \frac{\Theta_1 }{\Theta_1 - 1}    \bigg]  \bigg\}  \\ +  3   \underset{\Theta_1 : 1 > \frac{\Theta_1 - 1 }{\Theta_1 } > 0 }{\mathrm{sup}}  \bigg\{       1 -                 \textbf{P}_G  \bigg[  \forall \mathcal{G}  , \exists \bigg\{  \big\{ \mathscr{T}_{\textit{contraction}} \big\}  ,   \big\{  1 > \frac{\Theta_1 - 1 }{\Theta_1 }   > 0  \big\} \bigg\}    :         \mathscr{R}           =           \frac{\Theta_1 }{\Theta_1 - 1}    \bigg]        \bigg\} \\ +      \underset{\Theta_1 : 1 > \frac{\Theta_1 - 1 }{\Theta_1 } > 0 }{\mathrm{sup}}  \bigg\{                \textbf{P}_G  \bigg[  \forall \mathcal{G}  , \exists \bigg\{  \big\{ \mathscr{T}_{\textit{contraction}} \big\}  ,   \big\{  1 > \frac{\Theta_1 - 1 }{\Theta_1 }  > 0  \big\} \bigg\}   :    \frac{\Theta_1 - 1 }{\Theta_1}      <  \mathscr{R}         \bigg]              \bigg\}       \\ \\  \approx      1 +  3 \bigg[ 1 - \frac{1}{\epsilon} \bigg]     +    \underset{\Theta_1 : 1 > \frac{\Theta_1 - 1 }{\Theta_1 } > 0 }{\mathrm{sup}}  \bigg\{                       \textbf{P}_G  \bigg[  \forall \mathcal{G}  , \exists \bigg\{  \big\{ \mathscr{T}_{\textit{contraction}} \big\}  ,   \big\{  1 > \frac{\Theta_1 - 1 }{\Theta_1 }   > 0  \big\} \bigg\}    :         \mathscr{R}         \end{align*}

  \begin{align*}    =           \frac{\Theta_1 }{\Theta_1 - 1}    \bigg]        \bigg\}       \approx      1 + 3 \bigg[ 1 - \frac{1}{\epsilon} \bigg]  +   \frac{1}{\epsilon}  \approx 1                                                                                    ,
\end{align*} }

\noindent for $\epsilon$ taken sufficiently large, under the assumption that,

{\small \begin{align*}
 \big\{  P_1    +       P_2      +         P_3        \big\}   P^{-1}_4    > 1       , 
\end{align*} }

\noindent where,

{\small \begin{align*}
  P_1 \equiv  \textbf{P}_G  \bigg[  \forall \mathcal{G}  , \exists \bigg\{  \big\{ \mathscr{T}_{\textit{contraction}} \big\}  ,   \big\{  1 > \frac{\Theta_1 - 1 }{\Theta_1 }  > 0  \big\} \bigg\}   :    \frac{\Theta_1 - 1 }{\Theta_1}       <  \mathscr{R}           <          \frac{\Theta_1 }{\Theta_1 - 1}    \bigg]     ,  \\ \\ P_2 \equiv   \textbf{P}_G  \bigg[  \forall \mathcal{G}  , \exists \bigg\{  \big\{ \mathscr{T}_{\textit{contraction}} \big\}   ,   \big\{  1 > \frac{\Theta_1 - 1 }{\Theta_1 }   > 0  \big\} \bigg\}    :         \mathscr{R}           =           \frac{\Theta_1 }{\Theta_1 - 1}    \bigg]  , \\ \\ P_3 \equiv            \textbf{P}_G  \bigg[  \forall \mathcal{G}  , \exists \bigg\{  \big\{ \mathscr{T}_{\textit{contraction}} \big\}  ,   \big\{  1 > \frac{\Theta_1 - 1 }{\Theta_1 }  > 0  \big\} \bigg\}   :    \frac{\Theta_1 - 1 }{\Theta_1}      <  \mathscr{R}         \bigg] , \\ \\ P_4 \equiv   \textbf{P}_G  \bigg[  \forall \mathcal{G}  , \exists \bigg\{  \big\{ \mathscr{T}_{\textit{contraction}} \big\}  ,   \big\{  1 > \frac{\Theta_1 - 1 }{\Theta_1 }  > 0  \big\} \bigg\}   :    \frac{\Theta_1 - 1 }{\Theta_1}       <  \mathscr{R}           <          \frac{\Theta_1 }{\Theta_1 - 1}    \bigg] , 
\end{align*} }

\noindent from which we conclude the argument. \boxed{}

\subsection{Theorem $\textit{2}$}

\noindent \textit{Proof of Theorem 2}. To argue that,

\begin{align*}
  \textbf{P}_G      \big[ \mathcal{E}_2        \big]                \approx  1 , 
\end{align*}

\noindent where $\mathcal{E}_2$ is the event,

{\tiny \begin{align*}
      \bigg\{   \forall \mathcal{G}  , \exists \bigg\{ \big\{ \mathscr{T}_{\textit{contraction}} \big\} , \big\{   1 >  \epsilon_2 \neq \epsilon_1 > 0 \big\} \bigg\}    : \epsilon_2  <  \frac{\mathrm{sup} \big\{  \omega_q \big( \mathrm{Odd-Cycle} \big)^{\otimes 2 }  \big|_{\textit{tensors in } \mathscr{T}_{\textit{contraction}}}     -    \omega_q \big( \mathrm{Odd-Cycle} \big)^{\otimes 2} \big|_{\textit{tensors in } \mathscr{T}^{-1}_{\textit{contraction}}}    \big\} }{\omega_c \big( \mathrm{Odd-Cycle} \big)^{\otimes 2 }               }       \\    <    \frac{1}{\epsilon_2 }        \bigg\}      ,
\end{align*}}

\noindent directly apply the arguments from the previous result above, with,

{\tiny \begin{align*}
       \bigg\{   \forall \mathcal{G}  , \exists \bigg\{ \big\{ \mathscr{T}_{\textit{contraction}} \big\} , \big\{   1 >  \epsilon_2 \neq \epsilon_1 > 0 \big\} \bigg\}    : \epsilon_2  <  \frac{\mathrm{sup} \big\{  \omega_q \big( \mathrm{Odd-Cycle} \big)^{\otimes 2 }  \big|_{\textit{tensors in } \mathscr{T}_{\textit{contraction}}}     -    \omega_q \big( \mathrm{Odd-Cycle} \big)^{\otimes 2} \big|_{\textit{tensors in } \mathscr{T}^{-1}_{\textit{contraction}}}    \big\} }{\omega_c \big( \mathrm{Odd-Cycle} \big)^{\otimes 2 }               }              \\  <    \frac{1}{\epsilon_2 }        \bigg\}_{\epsilon_2 \leq \Theta_2 }   ,
\end{align*}}

\noindent and with,

{\tiny \begin{align*}
       \bigg\{   \forall \mathcal{G}  , \exists \bigg\{ \big\{ \mathscr{T}_{\textit{contraction}} \big\} , \big\{   1 >  \epsilon_2 \neq \epsilon_1 > 0 \big\} \bigg\}    : \epsilon_2  <  \frac{\mathrm{sup} \big\{  \omega_q \big( \mathrm{Odd-Cycle} \big)^{\otimes 2 }  \big|_{\textit{tensors in } \mathscr{T}_{\textit{contraction}}}     -    \omega_q \big( \mathrm{Odd-Cycle} \big)^{\otimes 2} \big|_{\textit{tensors in } \mathscr{T}^{-1}_{\textit{contraction}}}    \big\} }{\omega_c \big( \mathrm{Odd-Cycle} \big)^{\otimes 2 }               }             \\   \leq     \frac{1}{\epsilon_2 }        \bigg\}_{\epsilon_2 \leq \Theta_2 }     ,
\end{align*}}

\noindent from which we conclude the argument. \boxed{}

\subsection{Proposition}

\noindent \textit{Proof of Proposition}. To argue that,

{\small \begin{align*}
     \textbf{P}_G      \big[ \mathcal{E}_2        \big]  \textbf{P}_G     \bigg[  \forall \mathcal{G} , \exists  \bigg\{ \big\{ \textit{foam} \big\} , \big\{ \textit{tube} \big\}  , \big\{ \textit{section} \big\} , \big\{ m > 0 \big\} , \big\{ d \equiv 2 \big\}  \bigg\} : \big\{ \textit{(1)}  \big\} , \big\{       \textit{(2)}                \big\}     ,        \big\{      \textit{(3)}    \big\}     ,     \big\{     \textit{(4)}    \big\}  \text{ } \textit{simul-} \\ \textit{taneously occur}    \bigg]^{-1}       \\ \lesssim 1        ,
\end{align*}    }

\noindent implies,

{\small \begin{align*}
     \textbf{P}_G      \big[ \mathcal{E}_2        \big]  \textbf{P}_G     \bigg[  \forall \mathcal{G} , \exists  \bigg\{ \big\{ \textit{foam} \big\} , \big\{ \textit{tube} \big\}  , \big\{ \textit{section} \big\} , \big\{ m > 0 \big\} , \big\{ d \equiv 2 \big\}  \bigg\} : \big\{ \textit{(1)}  \big\} , \big\{       \textit{(2)}                \big\}     ,        \big\{      \textit{(3)}    \big\}     ,     \big\{     \textit{(4)}    \big\}  \text{ } \textit{simul-} \\ \textit{taneously occur}    \bigg]^{-1}   \\ \leq   \mathscr{F}_1     \mathscr{F}_2          \mathscr{F}^{-1}_3 \mathscr{F}_4  \mathscr{F}_5 \mathscr{F}^{-1}_6              ,
\end{align*}    }

\noindent from the prefactors $\mathscr{F}$, observe, by direct computation,

{\small\begin{align*}
   \textbf{P}_G      \big[ \mathcal{E}_2        \big]  \textbf{P}_G     \bigg[  \forall \mathcal{G} , \exists  \bigg\{ \big\{ \textit{foam} \big\} , \big\{ \textit{tube} \big\}  , \big\{ \textit{section} \big\} , \big\{ m > 0 \big\} , \big\{ d \equiv 2 \big\}  \bigg\} : \big\{ \textit{(1)}  \big\} , \big\{       \textit{(2)}                \big\}     ,        \big\{      \textit{(3)}    \big\}     ,     \big\{     \textit{(4)}    \big\}  \text{ } \textit{simul-} \\ \textit{taneously occur}    \bigg]^{-1}    
\end{align*}

\begin{align*}
 =   \textbf{P}_G     \bigg[    \forall \mathcal{G}  , \exists \bigg\{ \big\{ \mathscr{T}_{\textit{contraction}} \big\} , \big\{   1 >  \epsilon_2 \neq \epsilon_1 > 0 \big\} \bigg\}    : \epsilon_2  <  \frac{\mathrm{sup} \big\{ \omega_q \big( \mathrm{Odd-Cycle} \big)^{\otimes 2} \big\} }{\omega_q \big( \mathrm{Odd-Cycle} \big)^{\otimes 2}              \bigg|_{\textit{contraction}} }              <    \frac{1}{\epsilon_2 }        \bigg] \\ \times     \textbf{P}_G     \bigg[  \forall \mathcal{G} , \exists  \bigg\{ \big\{ \textit{foam} \big\} , \big\{ \textit{tube} \big\}  , \big\{ \textit{section} \big\} , \big\{ m > 0 \big\} , \big\{ d \equiv 2 \big\}  \bigg\} : \big\{ \textit{(1)}  \big\} , \big\{       \textit{(2)}                \big\}     ,        \big\{      \textit{(3)}    \big\}     ,     \big\{     \textit{(4)}    \big\}  \text{ } \textit{simult-} \\ \textit{aneously occur}    \bigg]^{-1}       \\ \\     \approx  \textbf{P}_G     \bigg[    \forall \mathcal{G}  , \exists \bigg\{ \big\{ \mathscr{T}_{\textit{contraction}} \big\} , \big\{   1 >  \epsilon_2 \neq \epsilon_1 > 0 \big\} \bigg\}    : \epsilon_2  <  \frac{\mathrm{sup} \big\{ \omega_q \big( \mathrm{Odd-Cycle} \big)^{\otimes 2} \big\} }{\omega_q \big( \mathrm{Odd-Cycle} \big)^{\otimes 2}              \bigg|_{\textit{contraction}} }              <    \frac{1}{\epsilon_2 }        \bigg]  \\ \times  {  \underset{1 \leq i \leq 4}{\prod} \textbf{P}_G     \bigg[  \forall \mathcal{G} , \exists  \bigg\{ \big\{ \textit{foam} \big\} , \big\{ \textit{tube} \big\}  , \big\{ \textit{section} \big\} , \big\{ m > 0 \big\} , \big\{ d \equiv 2 \big\}  \bigg\} : \big\{ \textit{(i)}  \big\}  \text{ } \textit{ occurs}    \bigg]^{-1}     }                \\ \\    \propto  \bigg|       \bigg\{            \textit{tensor}    :        \big\{ \textit{the image of tensor belongs to the image of $\mathscr{T}_{\textit{contraction}}$} \big\} ,    \big\{   \textit{the preimage oif tensor be-} \\ \textit{longs to the preimage of $\mathscr{T}_{\textit{contraction}}$}          \big\}    ,  \big\{       \textit{the tensor has support over $V \big( \mathcal{G} \big) \cap   V \big( \textit{tube} \big)    $}             \big\} , \big\{     \textit{the} \\ \textit{tensor has support over $V \big( \textbf{T}^2 \big) \cap V \big( \textit{tube} \big) $}       \big\}         \bigg\}                       \bigg|        .
\end{align*}
}

\noindent The number of tensors satisfying the conditions provided in the constant of proportionality in the last step above can approximately factored into the desired product of each factor, from which we conclude the argument. \boxed{}

\subsection{Theorem $\textit{3}$}

\noindent \textit{Proof of Theorem 3}. Denote,

{\small \begin{align*}
   1 \equiv   \bigg\{ \bigg| \frac{V \big( \textit{tube} \big)}{V \big( \mathcal{G} \big) } \bigg|  \lesssim \frac{1}{2} n^d \bigg\}  ,  \end{align*}

   \begin{align*}   2 \equiv \bigg\{  \bigg|  \frac{V \big( \textit{tube} \big)}{V \big( \textit{section} \big) }  \bigg|      \lesssim  \frac{1}{2} n^d \bigg\}   , \\ \\ 3 \equiv   \big\{  \big| v \in V \big( \textit{tube} \big) \cap V \big( \mathcal{G} \big) : \mathrm{span} \big\{  v \big\}  \lesssim n^d                 \big|       \big\}      , \end{align*}
   
   \begin{align*} 4 \equiv \bigg\{  \bigg|   \textit{tensor} : \textit{tensor is a strategy that Alice and Bob can use so that the winning} \\ \textit{probability satisfies, }        \epsilon_1 \\  <    \frac{\mathrm{sup} \big\{  \omega_q \big( \mathrm{Odd-Cycle} \big)  \big|_{\textit{tensors in } \mathscr{T}_{\textit{contraction}}}     -    \omega_q \big( \mathrm{Odd-Cycle} \big) \big|_{\textit{tensors in } \mathscr{T}^{-1}_{\textit{contraction}}}    \big\} }{\omega_c \big( \mathrm{Odd-Cycle} \big)              }           <    \frac{1}{\epsilon_1 }        \bigg|     \bigg\}           ,  
\end{align*} }

\noindent and also,

{\small 
\begin{align*} 
  1^{\prime} \equiv     \bigg\{ \bigg|    \bigg[ \frac{V \big( \textit{tube} \big)}{V \big( \mathcal{G} \big) } \bigg]  \bigg| \bigg| \bigg[ \frac{V \big( \textit{tube} \big)}{V \big( \textit{section} \big) }  \bigg]  \bigg|    \lesssim n^d \bigg\}             , \\ \\ 2^{\prime} \equiv    \big\{      \big|    v \in V \big( \textit{tube} \big)  \cap V \big( \mathcal{G} \big)  : \mathrm{span} \big\{  v \big\}     \big| \lesssim n^d     \big\}                 , \end{align*}

  \begin{align*} 3^{\prime} \equiv       \bigg\{ \bigg|  \textit{tensor} : \textit{tensor is a strategy that Alice and Bob can use so that the winning probability sati-} \\ \textit{sfies, }    \epsilon_1   <    \frac{\mathrm{sup} \big\{  \omega_q \big( \mathrm{Odd-Cycle} \big)  \big|_{\textit{tensors in } \mathscr{T}_{\textit{contraction}}}     -    \omega_q \big( \mathrm{Odd-Cycle} \big) \big|_{\textit{tensors in } \mathscr{T}^{-1}_{\textit{contraction}}}    \big\} }{\omega_c \big( \mathrm{Odd-Cycle} \big)              }   \\         <    \frac{1}{\epsilon_1 }        \bigg| \bigg\}      , \\ \\ 4^{\prime} \equiv             \bigg\{ \bigg|  \textit{tensor} : \textit{under one round of parallel repetition, the tensor is a strategy that Alice and Bob} \\ \textit{  can use so that the winning probability satisfies, }             \epsilon_2 \\  <    \frac{\mathrm{sup} \big\{  \omega_q \big( \mathrm{Odd-Cycle} \big)^{\otimes 2 }  \big|_{\textit{tensors in } \mathscr{T}_{\textit{contraction}}}     -    \omega_q \big( \mathrm{Odd-Cycle} \big)^{\otimes 2} \big|_{\textit{tensors in } \mathscr{T}^{-1}_{\textit{contraction}}}    \big\} }{\omega_c \big( \mathrm{Odd-Cycle} \big)^{\otimes 2 }               }           <    \frac{1}{\epsilon_2}       \bigg|      \bigg\}            .
\end{align*}

}

\noindent To argue that,

{\small \begin{align*}
       \textbf{P}_G      \big[ \mathcal{E}_2        \big]  \textbf{P}_G     \bigg[  \forall \mathcal{G} , \exists  \bigg\{ \big\{ \textit{foam} \big\} , \big\{ \textit{tube} \big\}  , \big\{ \textit{section} \big\} , \big\{ m > 0 \big\} , \big\{ d \equiv 2 \big\}  \bigg\} : \big\{ \textit{(1)}  \big\} , \big\{       \textit{(2)}                \big\}     ,        \big\{      \textit{(3)}    \big\}     ,     \big\{     \textit{(4)}    \big\}  \text{ } \textit{simul-} \\ \textit{taneously occur}    \bigg]^{-1}  \\  \approx 1 , 
\end{align*}}

\noindent holds, observe that the above ratio of probabilities is straightforwardly related to the fact that,

{\small \begin{align*}
 \textbf{P}_G \bigg[ \forall \mathcal{G} , \exists  \bigg\{ \big\{ \textit{foam} \big\} , \big\{ \textit{tube} \big\}  , \big\{ \textit{section} \big\} , \big\{ m > 0 \big\} , \big\{ d \equiv 2 \big\} , \big\{ 1 > \epsilon_1 > 0 \big\}  \bigg\}  :   1     , 2                   ,  3     ,   4                          \bigg] \\ \\ \approx   \textbf{P}_G \bigg[ \forall \mathcal{G} ,   \exists  \bigg\{ \big\{ \textit{foam} \big\} , \big\{ \textit{tube} \big\}  , \big\{ \textit{section} \big\} , \big\{ m > 0 \big\} , \big\{ d \equiv 2 \big\} ,  \big\{ 1 > \epsilon_2 \neq \epsilon_1 > 0 \big\}   \bigg\}  :     1^{\prime}   ,    2^{\prime}    ,    3^{\prime}  \\ ,  4^{\prime}                   \bigg]  .
\end{align*}} 

\noindent As $\big| V \big( \textit{tube} \big) \big| \longrightarrow \big|  V \big( \mathcal{G} \big) \big| $,

\begin{align*}
 \bigg|  \frac{V \big( \textit{tube} \big)}{V \big( \mathcal{G} \big) }  \bigg|  \bigg|  \frac{V \big( \textit{tube} \big)}{V \big( \textit{section} \big) }  \bigg|      \approx 1    , 
\end{align*}

\noindent from which we conclude the argument. \boxed{}

\section{Conclusion}

\noindent In this paper we argued that a problem for approximating the surface area of foam, which is related to similar problems considered in computational geometry, can also be characterized in the context of parallel repetition. In previous works of the author, parallel repetition were related to Alice and Bob \textit{approximately} obtaining the maximum winning probability. As a result of the correspondence between Odd Cycle elimination, and foam, problems, straightfowardly one can formulate each  probabilities depending upon the unmarked giant connected component, $\mathcal{G}$. From previous work in the literature, it has been assumed that properties $\mathcal{G}$ have helped past researchers establish conditions oo the typical number of connected components, and hence the total mass, of $\mathcal{G}$, over tubes of $\textbf{T}^2$. For the purposes of this effort, it is interesting to further examine $\mathcal{G}$ and its interactions with the optimal value. First, to quantify how the optimal value of the Odd-Cycle game is impacted by tensors that are removed in the contraction mapping which corresponds to the removal of cycles, we determine whether there exists a suitable constant that can be used to bound $  \frac{\mathrm{sup} \big\{  \omega_q \big( \mathrm{Odd-Cycle} \big)  \big|_{\textit{tensors in } \mathscr{T}_{\textit{contraction}}}     -    \omega_q \big( \mathrm{Odd-Cycle} \big) \big|_{\textit{tensors in } \mathscr{T}^{-1}_{\textit{contraction}}}    \big\} }{\omega_c \big( \mathrm{Odd-Cycle} \big)               }      $ from above, and from below. Second, to quantify how the optimal value of the Odd-Cycle game is impacted under one application of ordinary parallel repetition, we determine whether there exists a suitable constant that can be used to bound $  \frac{\mathrm{sup} \big\{  \omega_q \big( \mathrm{Odd-Cycle} \big)^{\otimes 2}  \big|_{\textit{tensors in } \mathscr{T}_{\textit{contraction}}}     -    \omega_q \big( \mathrm{Odd-Cycle} \big)^{\otimes 2} \big|_{\textit{tensors in } \mathscr{T}^{-1}_{\textit{contraction}}}    \big\} }{\omega_c \big( \mathrm{Odd-Cycle} \big)               }      $ from above, and from below. From the correspondence mentioned which relates Cycle elimination, Odd-Cycle elimination, and Foam problems to each other, one can straightforwardly establish connections between error inequalities for the Odd-Cycle game with computational problems associated with maximizing the surface area of foams.

\section{Declarations}

\subsection{Availability of data and materials}

Not applicable.

\subsection{Competing interests}

Not applicable.

\subsection{Funding}

Not applicable.

\subsection{Authors' contributions}

PR wrote the entire manuscript and performed several rounds of editing.

\subsection{Acknowledgments}

Not applicable.

\section{References}

\noindent [1] Azimian, K., Szegedy, M. (2008). Parallel Repetition of the Odd-Cycle Game. In: Laber, E.S., Bornstein, C., Nogueira, L.T., Faria, L. (eds) LATIN 2008: Theoretical Informatics. LATIN 2008. Lecture Notes in Computer Science, vol 4957. Springer, Berlin, Heidelberg. https://doi.org/10.1007/978-3-540-78773-0 58.

\bigskip

\noindent [2] Bannik, T. et al. Bounding Quantum-Classical Separations for Classes of Nonlocal Games. \textit{STACS} \textbf{12}: 1-12 (2019). https://doi.org/10.4230/LIPIcs.STACS.

\noindent 2019.12.

\bigskip

\noindent [3] Briet, J. Buhrman, H., Toner, B. A generalized Grothendieck inequality and entanglement in XOR games. \textit{Comm. Math. Phys.} \textbf{305}: 827-843 (2011). $\mathrm{https://doi.org/ 10.1007/s00220-011-1280-3}$.

\bigskip

\noindent [4] Broadbent, A., Methot, A.A. On the power of non-local boxes. \textit{Theoretical Computer Science} \textbf{358}: 3-14 (2006). $\mathrm{
https://doi.org/10.1016/j.tcs.2005.08.035}$.

\bigskip

\noindent [5] Brassard, G., Broadbent, A., Tapp, A. QuantumPseudo-Telepathy. \textit{Found. Phys.} \textbf{35}: 1877-1907 (2005). $\mathrm{https://philpapers.org/rec/BRAQP}$.

\bigskip

\noindent [6] Benedetti, M. and Coyle, B. and Fiorentini, M. and Lubasch, M. and Rosenkranz, M. Variational Inference with a QuantumComputer. \textit{Phys Rev Applied} 16: 044057 (2021) https://doi.org/10.1103/PhysRevApplied

\noindent .16.044057. 

\bigskip



\noindent [7] Bittel, L. and Kliesch, M. Training Variational QuantumAlgorithms is NP-Hard. \textit{Physical Review Letters} 127: 120502 (2021). https://doi.org/10.1103/PhysRevLett

\noindent .127.120502



\bigskip

\noindent [8] Catani, L. and Faleiro, R. and Emeriau,P.E. and Mansfield,S. and Pappa, A. Connecting XOR and XOR* games. \textit{Phys. Rev. A.} 109: 012427 (2024). https://doi.org/10.1103/PhysRevA.109.012427. 



\bigskip

\noindent [9] Chen, H. and Vives, M. and Metcalf, M. Parametric amplification of an optomechanical Quantuminterconnect. \textit{Physical Review Research} 4: 043119 (2022). https://doi.org/10.1103/PhysRevResearch.4.043119



\bigskip

\noindent [10] Cong, I. and Duan, L. Quantumdiscriminant analysis for dimensionality reduction and classification. \textit{New Journal of Physics} 18: 073011 (2016). https://doi.org/10.1088/1367-2630/18/7/073011. 


\bigskip

\noindent [11] Cleve, R., Hoyer, P., Toner, B., Watrous, J. Consequences and Limits of Nonlocal Strategies. \textit{19th IEEE Annual Conference on Computational Complexity Proceedings}: 236-249 (2004). $\mathrm{https://doi.org/10.1109}$$\mathrm{/CCC.2004.1313847}$.

\bigskip

\noindent [12] Culf, E., Mousavi, H., and Spirig, T. "Approximation Algorithms for Noncommutative CSPs," 2024 \textit{IEEE 65th Annual Symposium on Foundations of Computer Science (FOCS)}, 920-929 (2024). https://doi.org/ 10.1109/FOCS61266.

\noindent 2024.00061.

\bigskip

\noindent [13] Cui, D., Malavolta, G., Mehta, A., Natarajan, A., Paddock, C., Schmidt, S., Walter, M., Zhang, T. A Computational Tsireslson's Theorem for the Value of Compiled XOR games. \textit{arXiv: 2402.17301} (2024).

\bigskip

\noindent [14] Doherty, A.C., Liang, Y.C., Toner, B., Wehner, S. The QuantumMoment Problem and Bounds on Entangled Multi-Prover Games. \textit{23rd Annual IEEE Conference on Computational Complexity} \textbf{8}: 1093-0159/08 (2018).

\bigskip

\noindent [15] Drmota, P., Main, D., Ainley, E.M., Agrawal, A., Araneda, G., Nadlinger, Srinivas, R., Cabello, A.,  et al. Experimental QuantumAdvantage in the Odd-Cycle Game. \textit{Phys. Rev. Lett.} \textbf{134}: 070201 (2025).
https://doi.org/10.1103/PhysRevLett.

\noindent 134.070201.


\bigskip

\noindent [16] Ewe, W-B. and Koh, D. E. and Goh, S. T. and Chu, H-S and Png, C. E. Variational Quantum-Based Simulation of Waveguide Modes. \textit{IEEE Transactions on Microwave Theory and Techniques} 70 (5): 2517-2525 (2022). https://doi.org/10.1109/TMTT.2022.3151510.

\bigskip

\noindent [17] Emeriau, P-E., Howard, M. and Mansfield, S. QuantumAdvantage in Information Retrieval. \textit{PRX Quantum} \textbf{3}, 020307 (2022). https://doi.org/10.1103/PRXQuantum

\noindent .3.020307.



\bigskip

\noindent [18] Feige, U., Kindler, G., O'Donnell, R. Understanding Parallel Repetition Requires Understanding Foams. Computational Complexity Conference (CCC) 2007: Proceedings of the Twenty-Second Annual IEEE Conference on Computational Complexity: 179-192. $\mathrm{https://doi.org/10.11009/CCC.2007.39.}$.

\bigskip

\noindent [19] Garg, D. and Ikbal, S. and Srivastava, S.K. and Vishwakarma, H. and Karanam, H. and Subramaniam, L.V. QuantumEmbedding of Knowledge for Reasoning. \textit{Advance in Neural Information Processing Systems} 32 (2019). https://papers.nips.cc/paper files/paper/2019/hash/cb12d7f933e7d102c52231bf62b8a678-Abstract.html


\bigskip

\noindent [20] Genoni, M.G. and Tufarelli, T. Non-orthogonal bases for Quantummetrology. \textit{Journal of Physics A: Mathematical and Theoretical} 52: 43 (2019). https://doi.org/10.1088/1751-8121/ab3fe0.


\bigskip


\noindent [21] Gidi, J.A. and Candia, B. and Munoz-Moller, A.D. and Rojas, A. and Pereira, L. and Munoz, M. and Zambrano, L. and Delgado, A. Stochastic optimization algorithms for Quantumapplications. \textit{Phys.Rev.A} 108: 032409 (2023). https://doi.org/10.1103/PhysRevA.108.032409. 



\bigskip

\noindent [22] Givi, P. and Daley, A.J. and Mavriplis, D. and Malik, M. QuantumSpeedup for Aeroscience and Engineering. \textit{AIAA} 58:8 (2020). 

https://ntrs.nasa.gov/api/citations/20200003505/downloads/20200003505.pdf.


\bigskip

\noindent [23] Helton, J.W., Mousavi, H., Nezhadi, S.S. et al. Synchronous Values of Games. \textit{Ann. Henri Poincaré} \textbf{25}, 4357–4397 (2024). https://doi.org/10.1007/s00023-024-01426-1

\bigskip

\noindent [24] Hadiashar, S.B. and Nayak, A. and Sinha, P. Optimal lower bounds for QuantumLearning via Information Theory. \textit{IEEE Transactions on Information Theory} 70(3): 1876--1896 (2024). https://doi.org/10.1109/

\noindent TIT.2023.3324527. 



\bigskip

\noindent [25] Hur, T. and Kim, L. and Park, D.K. Quantumconvolutional neural network for classical data classification. \textit{QuantumMachine Intelligence} 4: 3 (2022). https://doi.org/10.1007/s42484-021-00061-x. 



\bigskip

\noindent [26] Holmes, Z. and Coble, N.J. and Sornborger, A.T. and Subasi, Y. On nonlinear transformations in Quantumcomputation. \textit{Phys. Rev. Research} 5: 013105 (2023). https://doi.org/10.1103/PhysRevResearch.5.013105. 



\bigskip

\noindent [27] Jing, H. and Wang, Y. and Li, Y. Data-Driven QuantumApproximate Optimization Algorithm for Cyber-Physical Power Systems. \textit{arXiv}: 2204.00738 (2022). https://doi.org/10.48550/arXiv.2204.00738.


\bigskip

\noindent [28] Junge, M., Palazuelos, C. On the power of Quantumentanglement in multipartite QuantumXOR games. \textit{Journal of the London Mathematical Society} \textbf{110}(5) (2024).

\bigskip

\noindent [29] Kubo, K. and Nakagawa, Y.O. and Endo, S. and Nagayama, S. Variational Quantumsimulations of stochastic differential equations. \textit{Physical Review A} 103: 052425 (2021). https://doi.org/10.1103/PhysRevA.

\noindent 103.052425.


\bigskip


\noindent [30] Kribs, D.W. A Quantumcomputing primer for operator theorists. \textit{Linear Algebra and its Applications} 400: 147-167 (2005). https://doi.org/10.48550/arXiv.math/

\noindent 0404553.


\bigskip

\noindent [31] Li, R. Y. and Di Felice, R. and Rohs, R. and Lidar, D.A. Quantumannealing versus classical machine learning applied to a simplied computational biology problem. \textit{NPJ QuantumInformation} 4: 14 (2008). https://doi.org/10.1038/s41534-018-0060-8. 


%

\bigskip

\noindent [32] Mahdian, M. and Yeganeh, H.D. Toward a Quantumcomputing algorithm to quantify classical and Quantumcorrelation of system states. \textit{QuantumInformation Processing} 20: 393 (2021). https://doi.org/10.1007/

\noindent s11128-021-03331-6.


\bigskip

\noindent [33] Maldonado, T.J. and Flick, J. and Krastanov, S. and Galda, A. Error rate reduction of single-qubit gates via noise-aware decomposition into native gates. \textit{Scientific Reports} 12: 6379 (2022). https://doi.org/10.1038

\noindent /s41598-022-10339-0.



\bigskip

\noindent [34] Manby, F.R. and Stella, M. and Goodpaster, J.D. and Miller, T.F. A Simple, Exact Density-Functional-Theory Embedding Scheme. \textit{Journal of Chemical Theory and Computation} 8 (8): 2564-2568 (2012). https://doi.org/10.1021/ct300544e.


\bigskip


\noindent [35] Mensa, S. and Sahin, E. and Tacchino, F. and Barkoutsos, P.K. and Tavernelli, I. QuantumMachine Learning Framework for Virtual Screening in Drug Discovery: a Prospective QuantumAdvantage. \textit{Mach. Learn.: Sci. Technol.} 4: 015023 (2023) https://doi.org/10.1088/2632-2153/acb900.


\bigskip


\noindent [36] Nan Sheng, H.M. and Govono, M. and Galli, G. QuantumEmbedding Theory for Strongly-Correlated States in Materials. \textit{J. Chem. Theory Comput.} 17 (4): 2116-2125 (2021). https://doi.org/10.1021/acs.jctc.

\noindent 0c01258. 


\bigskip

\noindent [37] Ostrev, D. The structure of nearly-optimal Quantumstrategies for the $\mathrm{CHSH(n)}$ XOR games. \textit{QuantumInformation and Computation} 16 (13-14): 1191-1211 (2016). https://doi.org/10.26421/QIC16.13-14-6.


\bigskip


\noindent [38] Paine, A.E. and Elfving, V.E. and Kyriienko, O. QuantumKernel Methods for Solving Differential Equations. \textit{Physical Review A} 107: 032428 (2023). https://doi.org/10.1103/PhysRevA.107.032428. 


\bigskip


\noindent [39] Paudel, H.P., Syamlal, M., Crawford, S.E., Lee, Y-L, Shugayev, R.A., Lu, P., Ohodnicki, P.R., Mollot, D., Duan, Y. \textit{QuantumComputing and Simulations for Energy Applications: Review and Perspective. ACS Eng. Au}: 3 151-196 (2022). $\mathrm{https://doi.org/10.1021/ac}$$\mathrm{sengineeringau.1c00033}$.


\bigskip


\noindent [40] Przhiyalkovskiy, Y.V. Quantumprocess in probability representation of Quantummechanics. \textit{Journal of Physics A: Mathematical and Theoretical} 55: 085301 (2022). https://doi.org/10.1088/1751-8121/ac4b15.


\bigskip

\bigskip

\noindent [41] Perc, M. Statistical physics of human cooperation. \textit{Physics Reports} \textbf{687}: 1-51 (2017). $\mathrm{https://papers.ssrn}$ $\mathrm{.com/sol3/papers.cfm?abstract_id=2972841}$.

\bigskip

\noindent [42] 
Oded, R. and Vidick, T. 2015. QuantumXOR Games. \textit{ACM Transactions on Computation Theory} ]\textbf{7}(4): Art. No. 15. https://doi.org/10.1145/2799560.

\bigskip

\noindent [43] Ravishankar Ramanathan, R., Augusiak, R., and Murta, G. Generalized XOR games with $d$ outcomes and the task of nonlocal computation. \textit{Phys. Rev. A} \textbf{93}, 022333 (2016). $\mathrm{https://doi.org/10.1103/PhysRevA}$ $\mathrm{.93.022333}$.

\bigskip

\noindent [44] Rigas, P. Optimal, and approximately optimal, Quantumstrategies for $\mathrm{XOR^{*}}$ and $\mathrm{FFL}$ games. \textit{arXiv: 2311.12887} (2023), submitted.

\bigskip


\noindent [45] Rigas, P. Quantumstrategies, error bounds, optimality, and duality gaps for multiplayer XOR, $\mathrm{XOR}^{*}$, compiled XOR, $\mathrm{XOR}^{*}$, and strong parallel repetiton of XOR, $\mathrm{XOR}^{*}$, and FFL games. 	arXiv:2505.06322, submitted (2025). $\mathrm{
https://doi.org/10.48550/arXiv.2505.06322
}$


\bigskip

\noindent [46] Roscika, M., Mazurek, P., Grudka, A., Horodecki, M. Generalized XOR non-locality games with graph description on a square lattice. \textit{Journal of Phys A: Math. Theor.} \textbf{53} 265302 (2020). $\mathrm{https://doi.org/10.1088/1751}$ $\mathrm{-8121/ab8f3e}$

\bigskip

\noindent [47] Slofstra, W.Lower bounds on the entanglement needed to play XOR non-local games \textit{Journal of Mathematical Physics} 52 (10): 102202 (2011). https://doi.org/10.1063/1.3652924.

\bigskip


\noindent [48] van Dam, W. and Sasaki, Y. QuantumAlgorithms for Problems in Number Theory, Algebraic Geometry, and Group Theory. \textit{Diversities in QuantumComputation and QuantumInformation}: 79-105 (2012). https://doi.org/10.1142/

\noindent 9789814425988$\mathrm{-}$0003.


\bigskip


\noindent [49] Wang, Y. and Krstic, P.S. Multistate Transition Dynamics by Strong Time-Dependent Perturbation in NISQ era. \textit{J. Phys.} Commun.7: 075004 (2023). https://doi.org/10.1088/2399-6528/ace67a.


\bigskip


\noindent [50] Zhao, L. and Zhao, Z. and Rebentrost, P. and Fitzsimons, J. Compiling basic linear algebra subroutines for Quantumcomputers,
 \textit{QuantumMachine Intelligence} 3: 21 (2021). https://doi.org/10.1007/s42484-021-00048-8.

\end{document}